\documentclass[12pt]{iopart}

\usepackage{graphicx,setspace}
\usepackage{dsfont}
\usepackage{iopams}
\usepackage{mathrsfs}
\usepackage{psfrag}
\usepackage{cases}

\begin{document}
\title{Some exact solutions for light beams}

\author{T G Philbin}
\address{Physics and Astronomy Department, University of Exeter,
Stocker Road, Exeter EX4 4QL, UK.}
\eads{\mailto{t.g.philbin@exeter.ac.uk}}

\begin{abstract}
We give an infinite class of exact analytical solutions for monochromatic light beams with strong focusing. As the solutions do not contain integrals, they are easy to explore compared with diffraction-theory results for strongly focused light. All monochromatic beams can be decomposed into two standing waves, each proportional to a Hilbert transform of the other. This means a beam can be built from any standing wave and our class is derived using this procedure. We give a visual overview of some of the beams, which reveals many interesting energy and field structures, including vortices in the energy flow, angular momentum in the propagation direction, and knotted field lines. We also show how the method can be used to design beams with an arbitrary focal shape.
\end{abstract}

\section{Introduction}
It is surprisingly challenging to find exact analytical solutions for monochromatic light beams. In many cases of interest the paraxial approximation is valid and then exact solutions are not required. 
Relatively simple and experimentally useful expressions can be obtained in the paraxial approximation, such as the Hermite-Gaussian and Laguerre-Gaussian beams (see~\cite{zha09}, for example). But beams with strong focusing, so-called ``non-paraxial" beams, are also of great theoretical and experimental interest. Exact beam solutions offer valuable insight into the remarkable behaviour of electromagnetic fields and energy near the focal region. Here we present an infinite class of exact analytical solutions for beams with very strong focusing. We show examples with different polarization properties and several examples with angular momentum in the propagation direction.

When we refer to exact analytical solutions, we mean of course in position space ($\bi{r}$-space). Exact analytical solutions for beams in wave-vector space ($\bi{k}$-space) are trivial to obtain: beams are given by functions on the hemisphere in $\bi{k}$-space of radius $k_0=\omega/c$ and $k_z>0$ (for propagation in the positive $z$-direction).\footnote{We only consider vacuum solutions valid in all (position) space and finite at infinity, so we exclude evanescent components.} Defining the beam in $\bi{k}$-space is a very useful method of generating exact solutions~\cite{sta81,she94,ber98,kis17} and it was used in~\cite{ber98}, for example, to explore the behaviour of vortices in a scalar beam. The drawback of this method is that it does not generally give an analytical solution in $\bi{r}$-space: in almost all cases the Fourier transform to $\bi{r}$-space must be done numerically at each point, which makes the solutions time-consuming to explore. The solutions presented here essentially amount to evaluating the Fourier integral from $\bi{k}$-space to $\bi{r}$-space analytically.

The inconvenience of numerical integrations also occurs in the most important method for studying light in a focus, which is based on diffraction theory~\cite{ign19,ign20,wol59,ric59}. Here diffraction integrals must be numerically evaluated at each point in space to obtain the fields. Nevertheless this formalism remains indispensable for the study of strong focusing given its power and versatility  (see~\cite{sta96,kar98,mon09,bli11,pan11} for example). 

A very interesting set of exact analytical wave solutions in $\bi{r}$-space has been obtained by the complex-sources method~\cite{des71,she98,ula00,she01,mit13,sap12,tag15}. Here one takes wave solutions created by monochromatic point sources and displaces a position coordinate of each point source by an imaginary constant. The resulting wave solutions can be \emph{source free}, and moreover they can have beam-like behaviour, particularly for parameters that do not give strong focusing. Unfortunately these solutions have plane-wave components that propagate opposite to the main wave direction~\cite{sap12} (i.e.\ the solutions do not live on a hemisphere in $\bi{k}$-space). Moreover the counter-propagating ``parasitic waves"~\cite{sap12} grow in amplitude as parameters are changed to increase focusing.

Recently an important class of exact beam solutions in $\bi{r}$-space was found by Lekner~\cite{lek16}. This work was based on calculating the exact expression in $\bi{r}$-space for a scalar beam that had been defined in $\bi{k}$-space by Stamnes~\cite{sta81}, who studied its focusing properties~\cite{sta81,sta}. Lekner's ``proto-beam"~\cite{lek16}, dubbed a ``perfect wave" by Stamnes~\cite{sta81}, is contained in the class of solutions presented here. The relation between our results and those in~\cite{lek16} is described in detail in Section~\ref{sec:standing}.

We seek exact beam solutions in $\bi{r}$-space that do not contain integrals. The key advantage of such solutions is that they are easy to explore, as field and energy plots can be  computed quickly (this speed-up compared to doing integrals at each point $\bi{r}$ is quantified in Section~\ref{sec:standing}). We first describe our method for scalar beams and look at some examples. It will then be straightforward to build electromagnetic beams from the scalar solutions. We give a pictorial survey of some of the remarkable field and energy structures to be found in the light beams. Finally we show how our approach can also be used to obtain a wide class of solutions containing integrals that must be evaluated numerically, but with the advantage of allowing some control over the shape of the beam in the focal region.

\section{Beams from a spherical standing wave}  \label{sec:standing}
Consider a monochromatic scalar wave $\mathrm{Re}[\phi(\bi{r})e^{-\rmi\omega t}]$ obeying the Helmholtz equation
\begin{equation}
\left( \nabla^2+k_0^2\right)\phi(\bi{r})=0, \qquad k_0=\frac{\omega}{c}.
\end{equation}
Every wave $\phi(\bi{r})$ can be written
\begin{equation}   \label{Fouier}
\phi(\bi{r})= \frac{1}{(2\pi)^3} \int \rmd\bi{k} \, f(\bi{k}) \delta(k-k_0) e^{\rmi\bi{k\cdot r}}
\end{equation}
and thus lives on a sphere of radius $k_0$ in $\bi{k}$-space. We define a beam as a wave for which $f(\bi{k})$ only has support on a hemisphere of the $k_0$-sphere, so that all its plane-wave components propagate in the same direction along a straight line in $\bi{r}$-space.

Our starting point is the spherical standing wave
\begin{equation}   \label{spherical}
\phi_R(\bi{r})=\frac{\sin(k_0 r)}{2\pi r},
\end{equation}
where the subscript $\scriptstyle R$ indicates that (\ref{spherical}) is real, as are all standing waves up to a constant complex factor. The standing wave (\ref{spherical}) is the retarded minus the advanced solution for a monochromatic point source, which gives a \emph{source-free} wave.  In $\bi{k}$-space all standing waves populate the $k_0$-sphere such that antipodes have equal amplitude but not necessarily equal phase; this is demonstrated for (\ref{spherical}) by writing it in the form (\ref{Fouier}):
\begin{equation}   \label{sphFouier}
\phi_R(\bi{r})= \frac{1}{(2\pi)^3} \int \rmd\bi{k} \, \frac{\pi}{k_0} \delta(k-k_0) e^{\rmi\bi{k\cdot r}}.
\end{equation}
This simply shows that the spherical standing wave is formed by coherent equal-amplitude plane waves converging on $\bi{r}=0$ from all directions. We can generate a beam from the standing wave (\ref{spherical}) by removing all $k_z<0$ plane-wave components. This is trivial to do in $\bi{k}$-space but it is difficult to evaluate the integral in (\ref{sphFouier}) with the restriction $k_z>0$, and our aim is to obtain an exact expression in $\bi{r}$-space. We find a way of evaluating the required integral by considering the problem more generally.

\begin{figure}[!htbp]
\begin{center} 
\includegraphics[width=15.5cm]{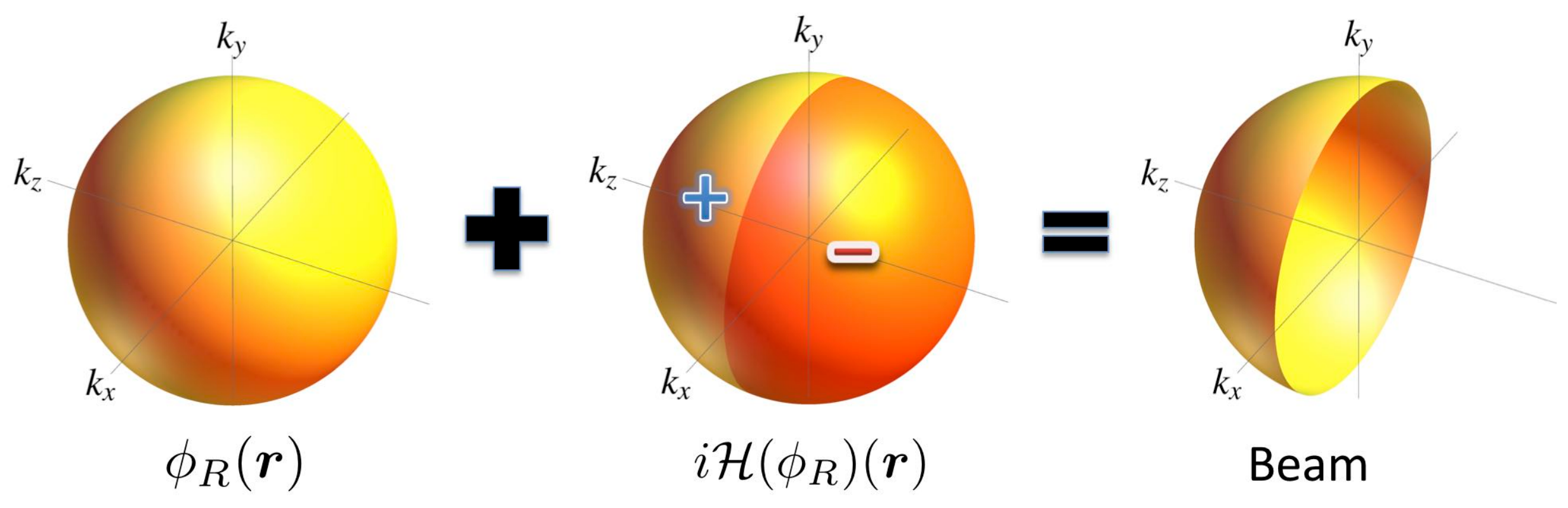}
\caption{Visualization in $\bi{k}$-space of the sum $\phi_R(\bi{r})+\rmi \mathcal{H}(\phi_R)(\bi{r})$ for the spherical standing wave (\ref{spherical}) and (\ref{sphFouier}). The spherical standing wave $\phi_R(\bi{r})$ populates equally all points on the $k_0$-sphere while $\rmi \mathcal{H}(\phi_R)(\bi{r})$ changes the sign of the $k_z<0$ plane-wave components. The sum is a beam that equally populates the $k_z>0$ hemisphere. }
\label{fig:hilbert}
\end{center}
\end{figure}

How do we perform in $\bi{r}$-space the operation of removing $k_z<0$ components from a standing wave $\phi_R(\bi{r})$? The answer is to add $\rmi \mathcal{H}(\phi_R)(\bi{r})$, where
\begin{equation}  \label{hilbert}
\mathcal{H}(\phi_R)(\bi{r}) :=\frac{1}{\pi} \mathrm{P}\int_{-\infty}^{\infty} \rmd s\,\frac{\phi_R(x,y,s)}{z-s}
\end{equation}
is a Hilbert transform in $z$. We can obtain this result in two ways. First, note from (\ref{Fouier}) that a beam in the positive $z$-direction is analytic in the upper-half complex $z$-plane because it only has $k_z>0$ plane-wave components. Now the imaginary part of a function that is analytic in the upper-half complex $z$-plane is the Hilbert transform of the real part~\cite{king}. Hence the function $\phi_R(\bi{r})+\rmi \mathcal{H}(\phi_R)(\bi{r})$ is analytic in the upper-half complex $z$-plane. Moreover $\mathcal{H}(\phi_R)(\bi{r})$ satisfies the Helmholtz equation because the Hilbert transform commutes with the derivatives $\nabla^2$. Therefore $\phi_R(\bi{r})+\rmi \mathcal{H}(\phi_R)(\bi{r})$ is a valid wave with only $k_z>0$ components, i.e.\ a beam.  The second way of seeing the result is to note that the Hilbert transform of the plane wave $e^{\rmi k_z z}$ is $-\rmi \,\mathrm{sign} (k_z) e^{\rmi k_z z}$. Hence when $\rmi$ times the Hilbert transform in $z$ of any wave (\ref{Fouier}) is added to the original wave, the $k_z<0$ components cancel, giving a beam in the positive $z$-direction. This latter description is visualized in Fig.~\ref{fig:hilbert}. The description in terms of plane-wave components also shows clearly that the Hilbert transform of any wave satisfies the Helmholtz equation and so gives another wave, as noted above. A trivial example that demonstrates the method is to start with the standing wave $\cos(k_0 z)$, take its Hilbert transform $\sin(k_0 z)$, and thereby obtain the beam $\cos(k_0 z)+\rmi \sin(k_0 z)=e^{\rmi k_0 z}$, i.e.\ a plane wave.

We can thus construct a beam from the spherical standing wave (\ref{spherical}) if we can find its Hilbert transform (\ref{hilbert}). Using the representation (\ref{sphFouier}) and the aforementioned fact that $\mathcal{H}(e^{\rmi k_z z})=-\rmi \,\mathrm{sign} (k_z) e^{\rmi k_z z}$, we obtain
\begin{equation}
\mathcal{H}(\phi_R)(\bi{r})=-\frac{\rmi \pi}{(2\pi)^3} \int \rmd\bi{k} \, \frac{\mathrm{sign}(k_z)}{k_0} \delta(k-k_0) e^{\rmi\bi{k\cdot r}}.
\end{equation}
Because of the $\mathrm{sign}(k_z)$ factor, the $k_z$-direction in $\bi{k}$-space is fixed by the $z$-direction in $\bi{r}$-space, but we are free to choose the  $k_x$-direction to lie along the direction $\boldsymbol{r_\perp} := x\boldsymbol{\hat{x}}+y\boldsymbol{\hat{y}}$.  Then transforming to spherical polar coordinates $\{k,\theta,\phi\}$ in $\bi{k}$-space we find
\begin{eqnarray}
\fl
\mathcal{H}(\phi_R)(\bi{r}) &= -\frac{\rmi \pi}{(2\pi)^3} \int_{0}^\pi \rmd \theta \int_0^{2\pi}\ \rmd\phi \, k_0 \sin\theta \, \mathrm{sign}(\cos\theta)  e^{\rmi k_0 \left( r_\perp  \sin\theta \cos\phi + z  \cos\theta \right)} \\
\fl
&=  -\frac{\rmi k_0}{ 4\pi} \int_{0}^\pi \rmd \theta  \sin\theta \, \mathrm{sign}(\cos\theta)  J_0\left(k_0 r_\perp \sin\theta \right)   e^{\rmi k_0  z  \cos\theta }  \\
\fl
&=  \frac{\rmi k_0}{ 4\pi}  \int_{1}^{-1} \rmd u \,  \mathrm{sign}(u)  J_0\left(k_0 r_\perp \sqrt{1-u^2} \right)   e^{\rmi k_0  z  u }  \\
\fl
&=  \frac{ k_0}{ 4\pi}  \int_{0}^{1} \rmd u \, J_0\left(k_0 r_\perp \sqrt{1-u^2} \right) \sin\left( k_0  z  u \right)   \label{Hsph0}  \\
\fl
 &=  - \sum_{n=0}^\infty\frac{ k_0 (-2)^{n-1} n! }{\pi(2 n+1)!} (k_0 z)^n
   \left(\frac{z}{r_\perp}\right)^{n+1}
   J_{n+1}\left(k_0 r_\perp\right)  .   \label{Hsph}
\end{eqnarray}
The step from (\ref{Hsph0}) to (\ref{Hsph}) is obtained by expanding the sine function as a power series and using ``Sonine's first finite integral" (see section~12.11 of~\cite{watson}, or integral 6.683(6) in~\cite{grad}). The expression (\ref{Hsph}) serves as the imaginary part of a beam whose real part is (\ref{spherical}), so the beam solution $\phi(\bi{r})=\phi_R(\bi{r})+\rmi \mathcal{H}(\phi_R)(\bi{r})$ is
\begin{equation}
\phi(\bi{r})=\frac{\sin(k_0 r)}{2\pi r}    - \rmi \sum_{n=0}^\infty\frac{ k_0 (-2)^{n-1} n! }{\pi(2 n+1)!} (k_0 z)^n
   \left(\frac{z}{r_\perp}\right)^{n+1}
   J_{n+1}\left(k_0 r_\perp\right)  .   \label{phibeam}
\end{equation}

The beam (\ref{phibeam}) is composed of plane waves converging on $\bi{r}=0$ from all directions lying on a hemisphere. It thus gives one representation of a maximally focusing beam. Although there is an infinite series in (\ref{phibeam}), which must be numerically summed to a sufficient degree to obtain the field at a point $\bi{r}$, the beam (\ref{phibeam}) is very easy to work with compared to solutions containing integrals. The number of terms in the sum which contribute significantly increases with distance from the focus $\bi{r}=0$ but in practice numerical evaluation is very fast. In almost all of the plots which follow, which were prepared using Mathematica, thirty terms were taken in the sum as this was more than sufficient for numerical accuracy. With the software used, the speed-up conferred by using (\ref{phibeam}) compared to evaluating the Fourier integral for (\ref{phibeam}) is around two orders of magnitude for the points $\bi{r}$ in the plots (the Fourier integral for (\ref{phibeam}) is the integral (\ref{stamnes}) below).

From (\ref{phibeam}) we in fact obtain an infinite class of exact beam solutions, since acting on (\ref{phibeam}) with spatial derivative operators produces other solutions. The $z$-derivative of (\ref{phibeam}) gives a beam that was defined in $\bi{k}$-space by Stamnes~\cite{sta81} and recently evaluated in $\bi{r}$-space by Lekner~\cite{lek16}. To show explicitly the relation to~\cite{sta81,lek16}, restrict the integral in (\ref{sphFouier}) to $k_z>0$ and expand the delta function in terms of $\delta(k_z-\sqrt{k_0^2-k_\perp^2})$. This gives the integral
\begin{equation}  \label{stamnes}
 \frac{2}{(2\pi)^3} \int_{k_\perp\leq k_0} \rmd\bi{k}_\perp \, \frac{\pi}{\sqrt{k_0^2-k_\perp^2}} \exp\left[\rmi \left(\bi{k}_\perp\bi{\cdot r}\!_\perp + \sqrt{k_0^2-k_\perp^2} \,z\right)\right],
\end{equation}
where we have inserted a factor of two because our procedure for removing the negative $k_z$ components of (\ref{sphFouier}) gives twice the positive $k_z$ components. We have proved that (\ref{stamnes}) evaluates to the beam (\ref{phibeam}). The $z$-derivative of (\ref{stamnes}) removes the denominator in the integrand and gives the beam defined in~\cite{sta81} and  evaluated in $\bi{r}$-space in~\cite{lek16}. The ``proto-beam" of~\cite{lek16} is thus the $z$-derivative of the beam (\ref{phibeam}) (apart from an imaginary constant factor that swaps the real and imaginary parts). Lekner indeed obtains the $z$-derivative of the spherical standing wave as one part of his proto-beam and finds an expression for the other part (which is the $z$-derivative of the infinite series in (\ref{phibeam})) as a \emph{finite} sum of terms involving Bessel and Lommel functions~\cite{lek16}. It is possible that the imaginary part of our beam (\ref{phibeam}) could also be expressed as a finite sum of special functions but the result would probably be very cumbersome to write down. It is easy to see from the $z$-derivative of (\ref{phibeam}) that the expression for the proto-beam in the focal plane $z=0$ is proportional to $J_{1}\left(k_0 r_\perp\right)/ r_\perp$, which is a feature of the proto-beam discussed in~\cite{sta81,sta,lek16}. The proto-beam was found in~\cite{and18} to meet a criterion for largest intensity increase from focusing.  We will discuss examples based on the basic beam solution (\ref{phibeam}) and derivatives thereof, but none of our examples will have a $z$-derivative of (\ref{phibeam}) and so they will not be in Lekner's class of solutions. We refer the reader to~\cite{lek16,and18,and17} for many interesting details of the proto-beam and electromagnetic beams built from it.

Lekner~\cite{lek16} also points out that further solutions can be generated by the method of displacing $z$ by a complex constant, which was used to generate the "complex-sources" waves described in the Introduction. This allows the class of beams in~\cite{lek16} and the class given here to be broadened further, but our examples below will not exploit this possibility.

In summary, we see that (\ref{phibeam}) is the fundamental solution in the infinite class, from which the $\bi{r}$-space expressions for all the other solutions in the class are easily obtained. The class of solutions includes beams with angular momentum in the direction of propagation, an example of which will be discussed in the next section.

\section{Scalar beams}   \label{sec:scal}
A comprehensive analysis of any of the scalar beam solutions of the last section would be quite lengthy. Here we briefly  show some interesting features of two of the beams.  

\begin{figure}[!htbp]
\begin{center} 
\includegraphics[width=15cm]{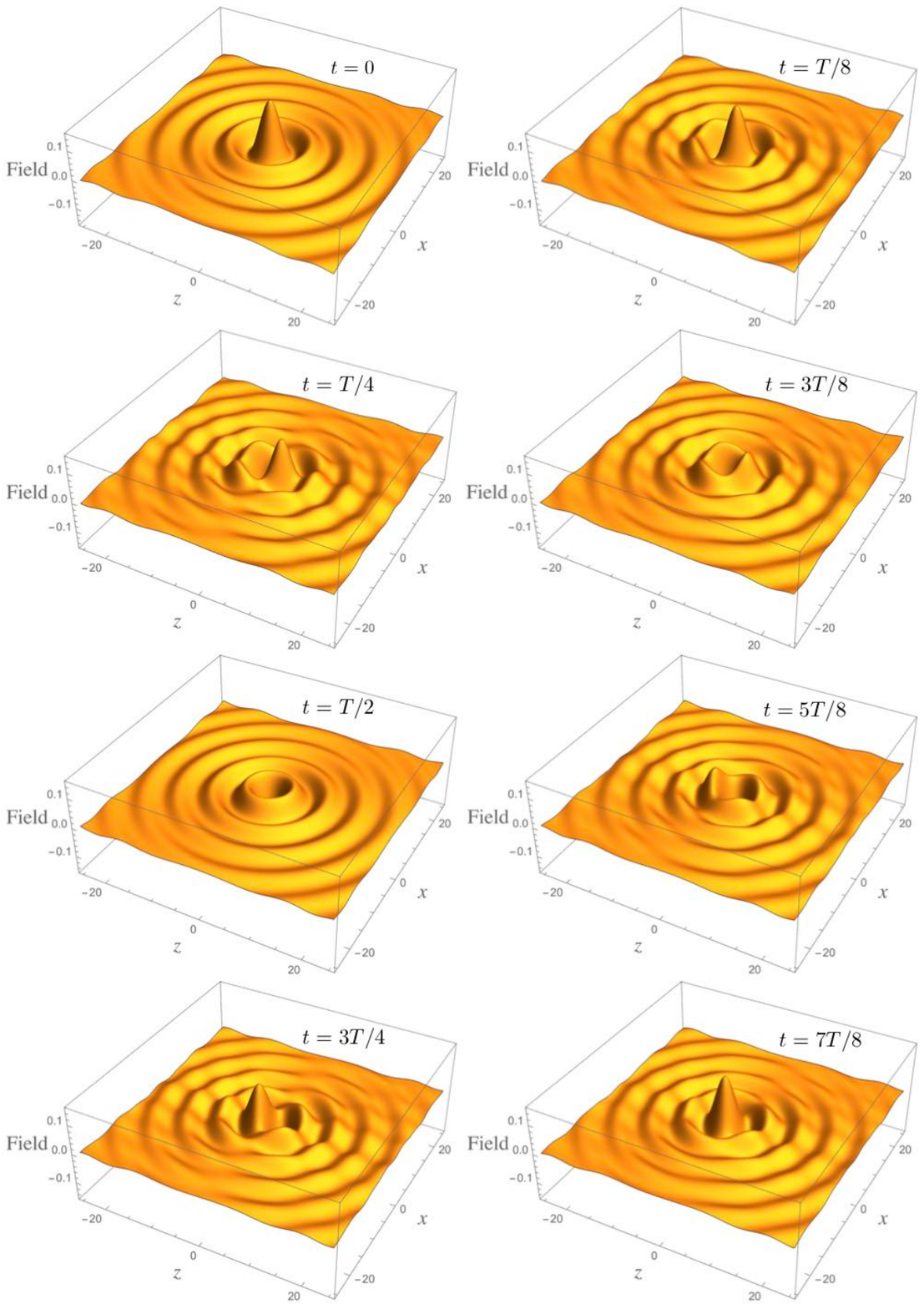}
\caption{ The field $\mathrm{Re}[\phi(\bi{r})e^{-\rmi\omega t}]$ of the beam (\ref{phibeam}) in the $zx$-plane at equal time intervals during one period $T=2\pi/\omega$. The field is rotationally invariant around the beam axis ($z$-axis). The parameters are $k_0=c=\omega=1$. }
\label{fig:scalarfield}
\end{center}
\end{figure}

Consider first the fundamental solution (\ref{phibeam}). The field $\mathrm{Re}[\phi(\bi{r})e^{-\rmi\omega t}]$ of this beam is plotted in Fig.~\ref{fig:scalarfield} at times separated by an eighth of a period $T=2\pi/\omega$. Twice during each period the field coincides with that of the spherical standing wave, for example at $t=0$ and $t=T/2$ in Fig.~\ref{fig:scalarfield}. The real (imaginary) part of (\ref{phibeam}) is symmetric (antisymmetric) in $z$ so the field in the focal plane $z=0$ is that of the spherical standing wave.\footnote{This should be contrasted with the field in the focal plane for the proto-beam, which as noted in Sec.~\ref{sec:standing} is proportional to  $J_{1}\left(k_0 r_\perp\right)/ r_\perp$~\cite{sta81,sta,lek16}. The proto-beam thus has a tighter focus in the focal plane.}  It therefore follows from (\ref{spherical}) that the beam has concentric rings of field nodes in the focal plane at $r_\perp=\pi n/k_0$, $n\in\mathds{Z}-\{0\}$. These nodes are centres of vortices in the focal plane (see below). Along the beam axis ($z$-axis) we can sum the infinite series in (\ref{phibeam}) using $J_{n+1}(\epsilon)= 2^{-n-1}\epsilon^{n+1}/(n+1)!+\Or(\epsilon^{n+2})$ and obtain
\begin{equation}  \label{scalzaxis}
\phi(0,0,z)=\frac{1}{2\pi z}\left[\sin\left( k_0 z \right) +2\rmi \sin^2\left(\frac{1}{2} k_0 z \right)    \right]  =  -\frac{\rmi}{2\pi z}\left( e^{\rmi k_0 z} -1   \right)  .
\end{equation}
This 1D wave is just $\frac{1}{2\pi}\int_0^{k_0} dk_z\, e^{\rmi k_z z}$, the sum of the $k_z\in[0,k_0]$ plane-wave components on the $k_0$-sphere. Note that $\phi(0,0,z)$ has nodes at $z=2\pi n/k_0$, $n\in\mathds{Z}-\{0\}$, which are not the locations of vortices (see below).

\begin{figure}[!htbp]
\begin{center} 
\includegraphics[width=14cm]{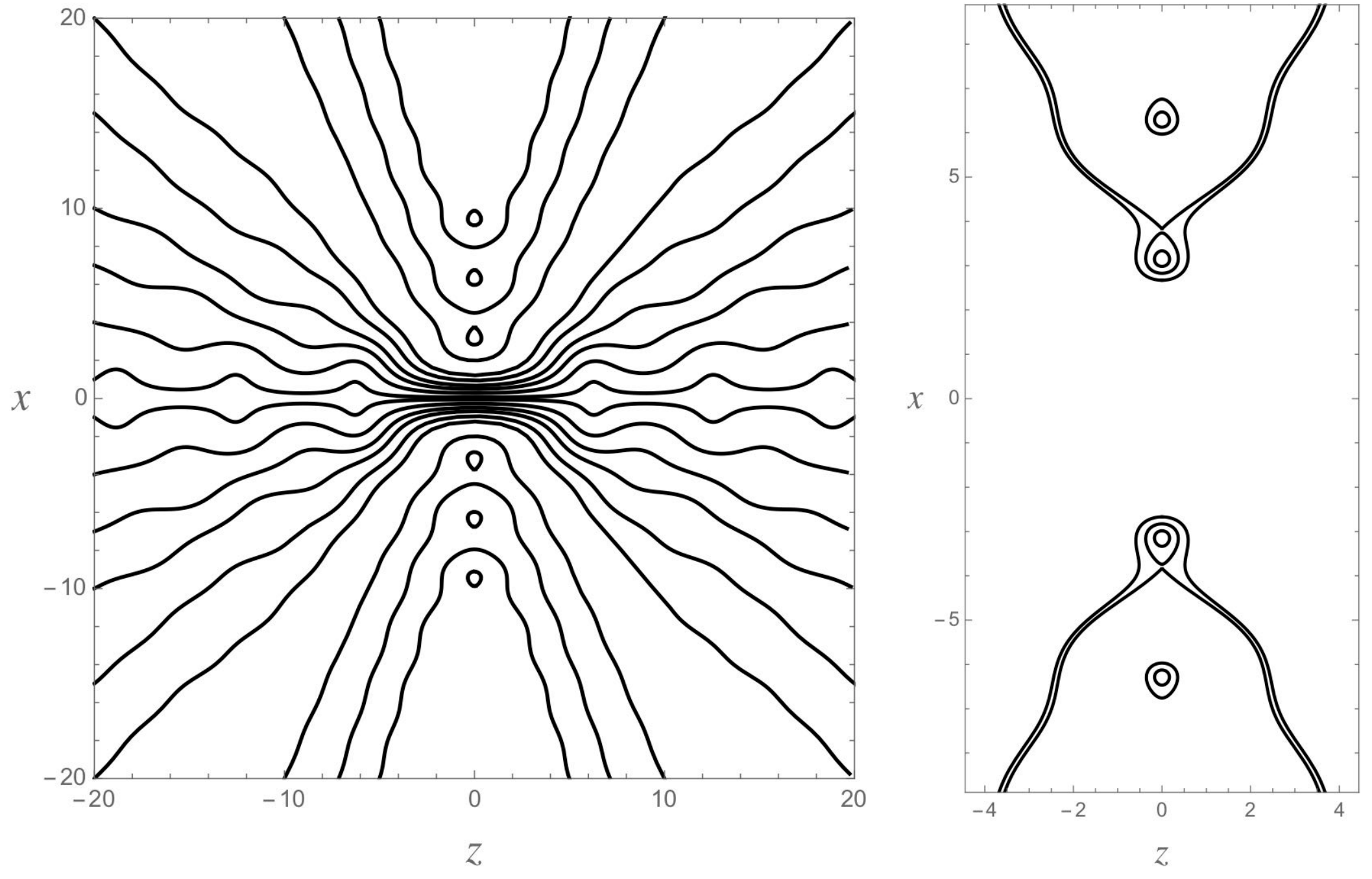}
\caption{ Integral curves of the time-averaged energy flux (\ref{Savscal}) for the beam (\ref{phibeam}) in the $xz$-plane. Vortices are visible in the region near the focal plane $z=0$; these are centred on circular nodes of the field in the $xy$-plane. Visible in the right-hand plot are phase saddle points where integral curves meet at a zero of the vector $\langle\bi{S}\rangle $. The parameters are $k_0=c=\omega=1$. }
\label{fig:Savscal}
\end{center}
\end{figure}

The energy flux vector (Poynting vector) of a scalar wave $\psi(\bi{r},t)$ is $\bi{S}=-[\boldsymbol{\nabla}\psi(\bi{r},t)] \partial_t\psi(\bi{r},t)$. For monochromatic waves $\mathrm{Re}[\phi(\bi{r})e^{-\rmi\omega t}]$ we have for  $\bi{S}$ and its time average  $\langle\bi{S}\rangle$:
\begin{eqnarray}
\fl
\bi{S} = \omega \left[  \phi_R(\bi{r}) \sin(\omega t) -  \phi_I(\bi{r} ) \cos(\omega t) \right] \left[ \boldsymbol{\nabla} \phi_I(\bi{r}) \sin(\omega t) + \boldsymbol{\nabla} \phi_R(\bi{r} ) \cos(\omega t) \right], \label{Sscal} \\
\fl
\langle\bi{S}\rangle  = \frac{1}{2} \left[  \phi_R(\bi{r})  \boldsymbol{\nabla} \phi_I(\bi{r} )   -  \phi_I(\bi{r} )  \boldsymbol{\nabla} \phi_R(\bi{r} ) \right],  \label{Savscal}
\end{eqnarray}
where $\scriptstyle R$ ($\scriptstyle I$) denotes real (imaginary) part. Integral curves of the time-averaged energy flux (\ref{Savscal}) for the beam (\ref{phibeam}) in a plane through the beam axis are shown in Fig.~\ref{fig:Savscal}. They show familiar features of strongly focused beams~\cite{born,den09,gbur}. There are vortices around the circular field nodes in the focal plane ($xy$-plane). Both $\langle\bi{S}\rangle $ and $\bi{S}$ are zero at all field nodes. The time-averaged energy flow lines close to the beam axis undulate to avoid the nodes located on the axis.  Associated with each vortex ring there is a ring of phase saddle points of $\phi(\bi{r})$~\cite{den09,gbur} where  $\langle\bi{S}\rangle $, but not $\bi{S}$, vanishes. Integral curves of $\langle\bi{S}\rangle $ meet at the phase saddle points (see Fig.~\ref{fig:Savscal}).

\begin{figure}[!htbp]
\begin{center} 
\includegraphics[width=9cm]{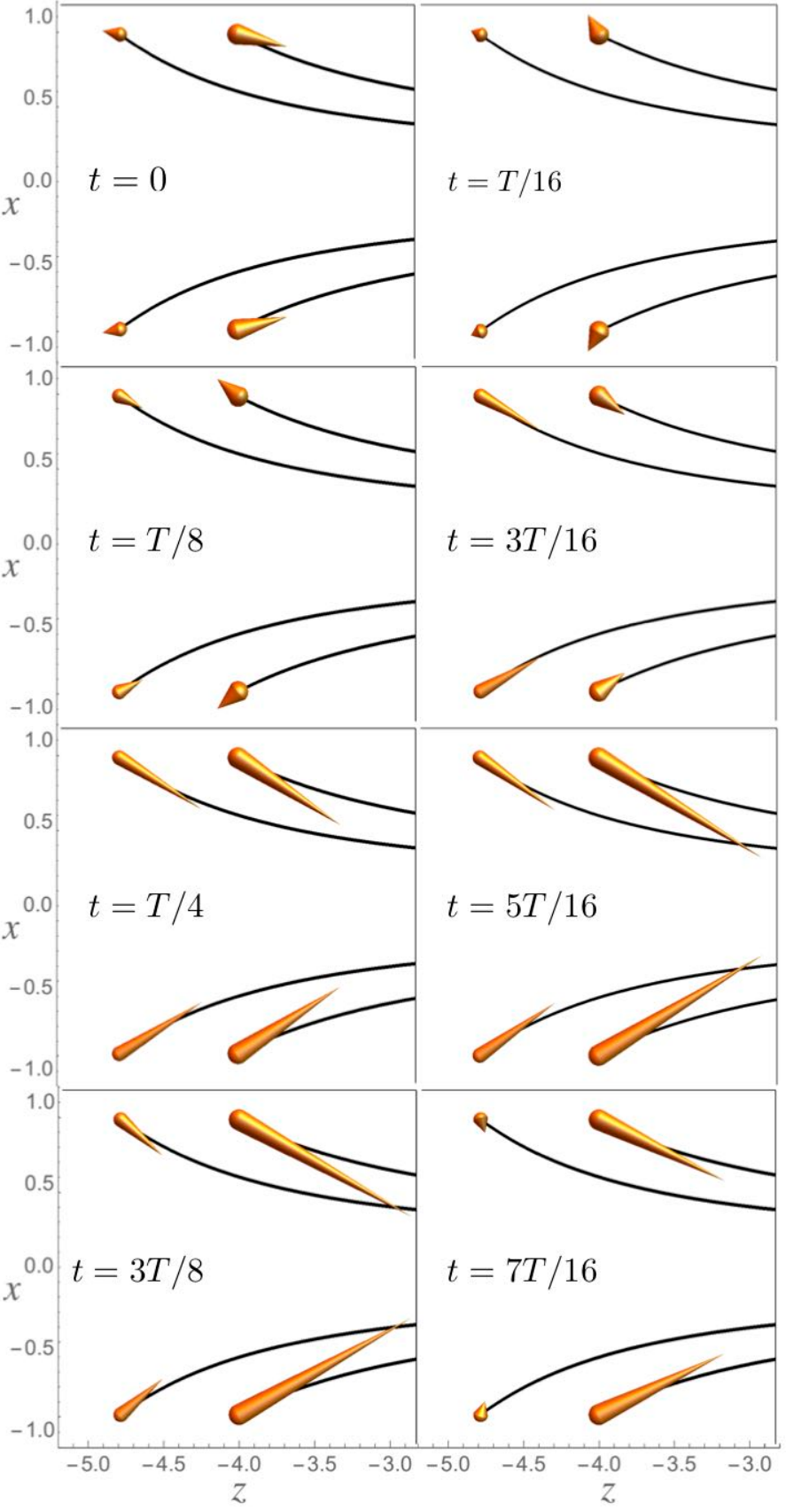}
\caption{ Energy flux $\bi{S}$ (conical gold pointers) at four points near the beam focus $\bi{r}=0$, at different times over a half-period of the beam. The black curves are integral curves of  $\langle\bi{S}\rangle$, so the time average of the gold pointers is tangent to these curves. Note that $\bi{S}$ is periodically opposite to the beam direction. The parameters are $k_0=c=\omega=1$. }
\label{fig:Sscal}
\end{center}
\end{figure}

It is interesting to compare $\langle\bi{S}\rangle $ with $\bi{S}$. Figure~\ref{fig:Sscal} shows $\langle\bi{S}\rangle $ and $\bi{S}$ at four points near the beam focus $\bi{r}=0$, together with the four integral curves of $\langle\bi{S}\rangle $ that pass through those points. The energy flux $\bi{S}$ cycles at half the field period $T$. Note that the energy flux at the four points is periodically opposite to the beam direction, even though the time-averaged flux is in the beam direction. The periodic flow of energy opposite to the beam direction gradually disappears as one moves to points further from the focus. This behaviour should be contrasted with that in vortex regions, where $\langle\bi{S}\rangle $ can be opposite to the beam direction (see Fig.~\ref{fig:Savscal}).

\begin{figure}[!htbp]
\begin{center} 
\includegraphics[width=15.5cm]{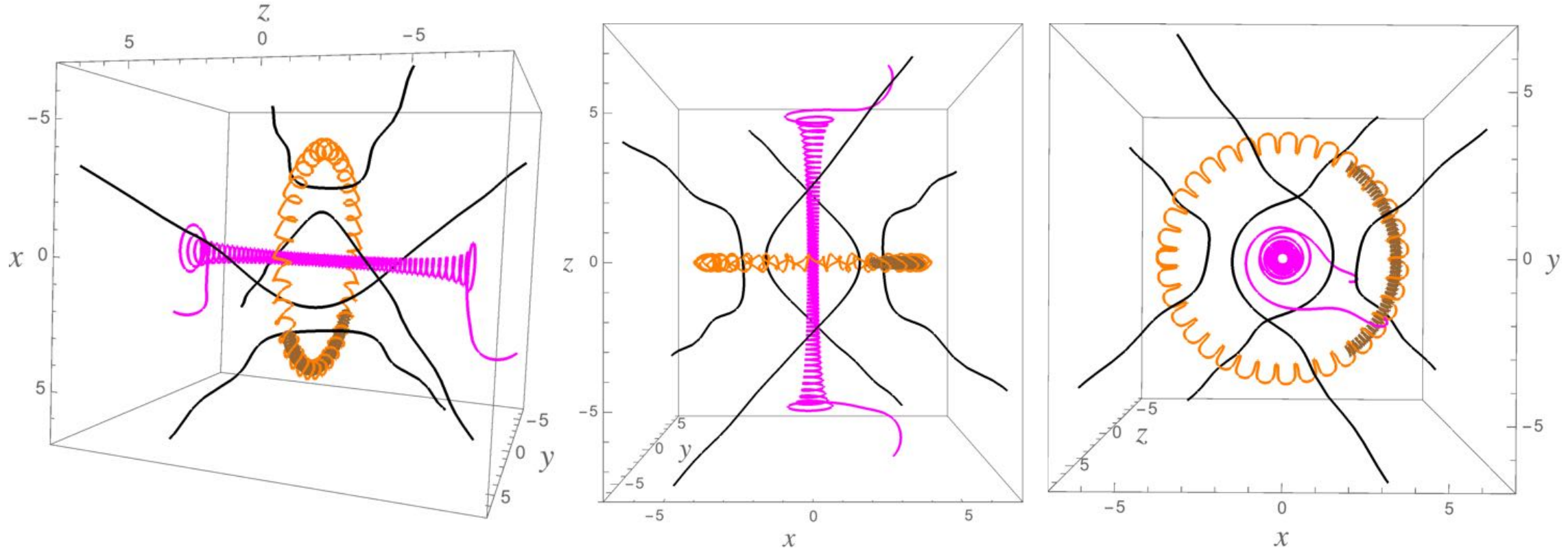}
\caption{Integral curves of the time-averaged energy flux (\ref{Savscal}) for the beam with OAM obtained by operating with $\partial_x+\rmi \partial_y$  on the beam (\ref{phibeam}). Three views of the same integral curves are shown. The vortex region closest to the beam axis is visible. Vortex regions lie close to circles of field nodes in the focal plane $z=0$. One complete closed loop of time-averaged energy flux is shown in the vortex region, and also part of a loop that lies closer to the circular field node (this partial loop is in the positive $x$ region of the plots). The latter loop winds more tightly around the circular node. The parameters are $k_0=c=\omega=1$. }
\label{fig:SavscalOAM}
\end{center}
\end{figure}

As a second example, we consider a scalar beam with angular momentum in  the propagation direction, often called orbital angular momentum (OAM). The simplest member of our class of beams that has nonzero OAM is obtained by acting with $\partial_x+\rmi \partial_y$ on the fundamental solution  (\ref{phibeam}). To motivate this, note that the operator $\partial_x+\rmi \partial_y$ turns the Bessel beam~\cite{bessel} $e^{\rmi k_0 z \cos\alpha } J_0(k_0 r_\perp \sin\alpha )$, which does not have OAM, into a Bessel beam with OAM, namely $e^{\rmi \phi+\rmi k_0 z \cos\alpha } J_1(k_0 r_\perp \sin\alpha ) $, where $\phi$ is the angle of cylindrical coordinates. The effect of $\partial_x+\rmi \partial_y$ on the  fundamental solution  (\ref{phibeam}) is not as simple, but in the focal plane $z=0$ it generates a factor $e^{\rmi \phi}$ and the resulting beam has OAM. We can see the OAM in the beam by plotting integral curves of the time-averaged energy flux $\langle\bi{S}\rangle $, see Fig.~\ref{fig:SavscalOAM}. The momentum density is proportional to the enegy flux and the beam has angular momentum in the direction of propagation (OAM) when $\bi{S}$ has a non-zero azimuthal component in planes perpendicular to the beam direction. The OAM is clearly visible in Fig.~\ref{fig:SavscalOAM}, which also shows a very tight twisting of the time-averaged energy flux around the beam axis  in the region close to the axis. The beam has vortices and Fig.~\ref{fig:SavscalOAM} contains the vortex region closest to the beam axis. As in the fundamental solution (\ref{phibeam}) (see Fig.~\ref{fig:Savscal}), vortex regions are toroidally shaped with circular field nodes (in the focal plane) at their centres. But whereas for the beam (\ref{phibeam}) the closed loops of  time-averaged energy flux lie in planes through the beam axis, in the beam with OAM the loops move around the beam axis while winding around the circular node. The number of windings of the loop around the circular node increases for loops closer to the node.

\section{Electromagnetic beams}  \label{sec:embeams}
Our main interest is light beams. We obtain an infinite class of light beams by employing Lorenz gauge and taking the scalar beams from the last section as Cartesian components of the vector potential. In Lorenz gauge the monochromatic electric field $\bi{E}(\bi{r},t)=\mathrm{Re}[\bi{E}_0(\bi{r})e^{-\rmi\omega t}]$ and magnetic field $\bi{B}(\bi{r},t)=\mathrm{Re}[\bi{B}_0(\bi{r})e^{-\rmi\omega t}]$ are given by~\cite{jac}
\begin{equation}   \label{EBA}
\bi{B}_0(\bi{r})=\boldsymbol{\nabla \! \times \! A}_0(\bi{r}), \qquad  \bi{E}_0(\bi{r})= i\omega \bi{A}_0(\bi{r})+i\frac{c^2}{\omega}\boldsymbol{\nabla} [\boldsymbol{\nabla \! \cdot \! A}_0(\bi{r})],
\end{equation}
where $\bi{A}(\bi{r},t)=\mathrm{Re}[\bi{A}_0(\bi{r})e^{-\rmi\omega t}]$ is the vector potential. In the Lorenz gauge each Cartesian component of $\bi{A}_0(\bi{r})$ obeys the Helmholtz equation if there are no sources~\cite{jac} and so we may use the scalar beams for these components. In the remainder of this section we give a pictorial survey of some the resulting light beams.

\begin{figure}[!htbp]
\begin{center} 
\includegraphics[width=14cm]{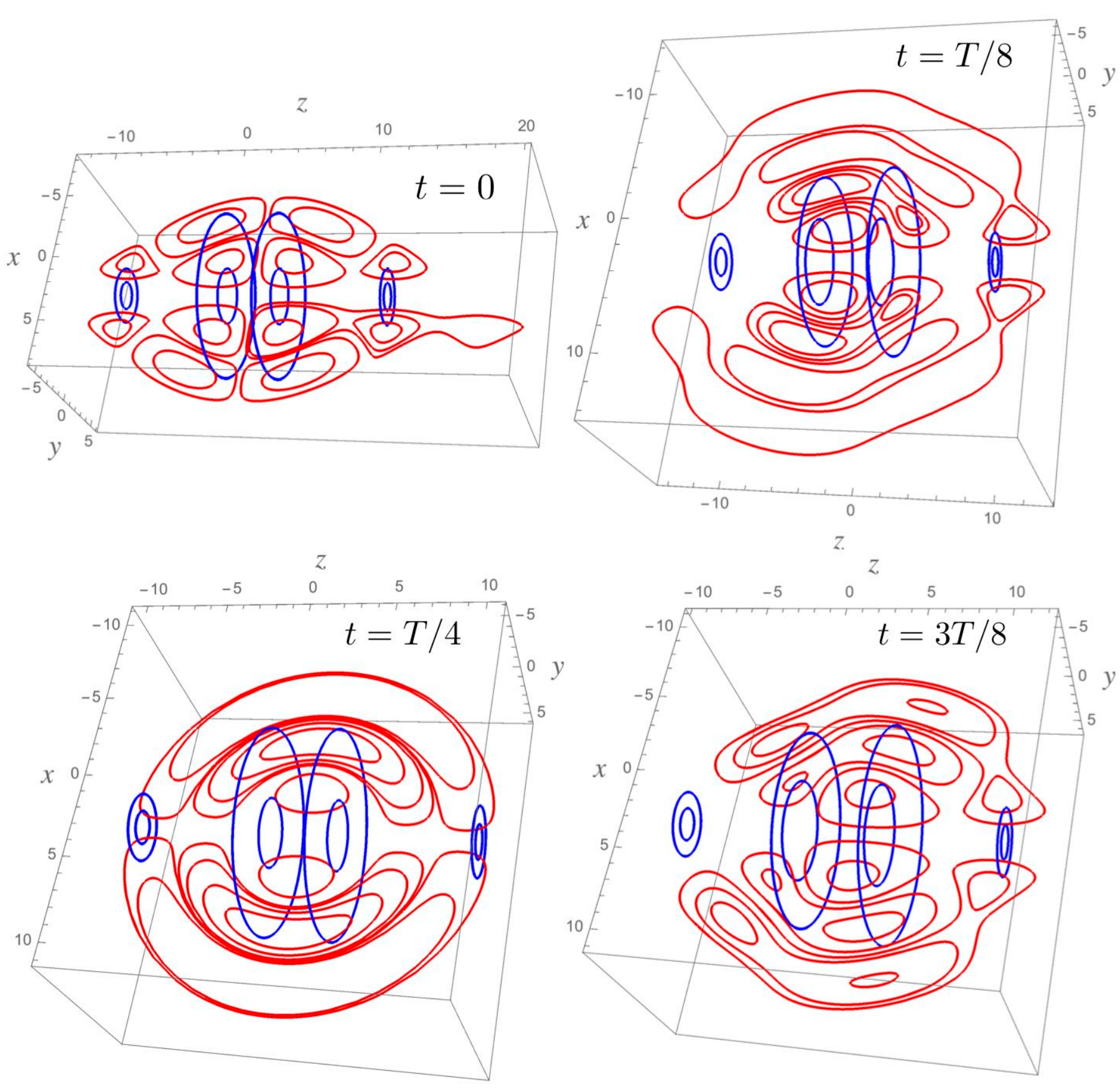}
\caption{Electric (red) and magnetic (blue) field lines (integral curves of $\bi{E}$ and $\bi{B}$) in the radially polarized beam of Sec.~\ref{sec:rad} at four times over a half-period $T/2$. The field lines repeat with opposite directions every half period. The $\bi{E}$ field is shown in the $xy$-plane and is identical in every plane through the beam axis. The parameters are $\mu_0=c=k_0=\omega=p_0=1$. }
\label{fig:fieldsrad}
\end{center}
\end{figure}
\begin{figure}[!htbp]
\begin{center} 
\includegraphics[width=10cm]{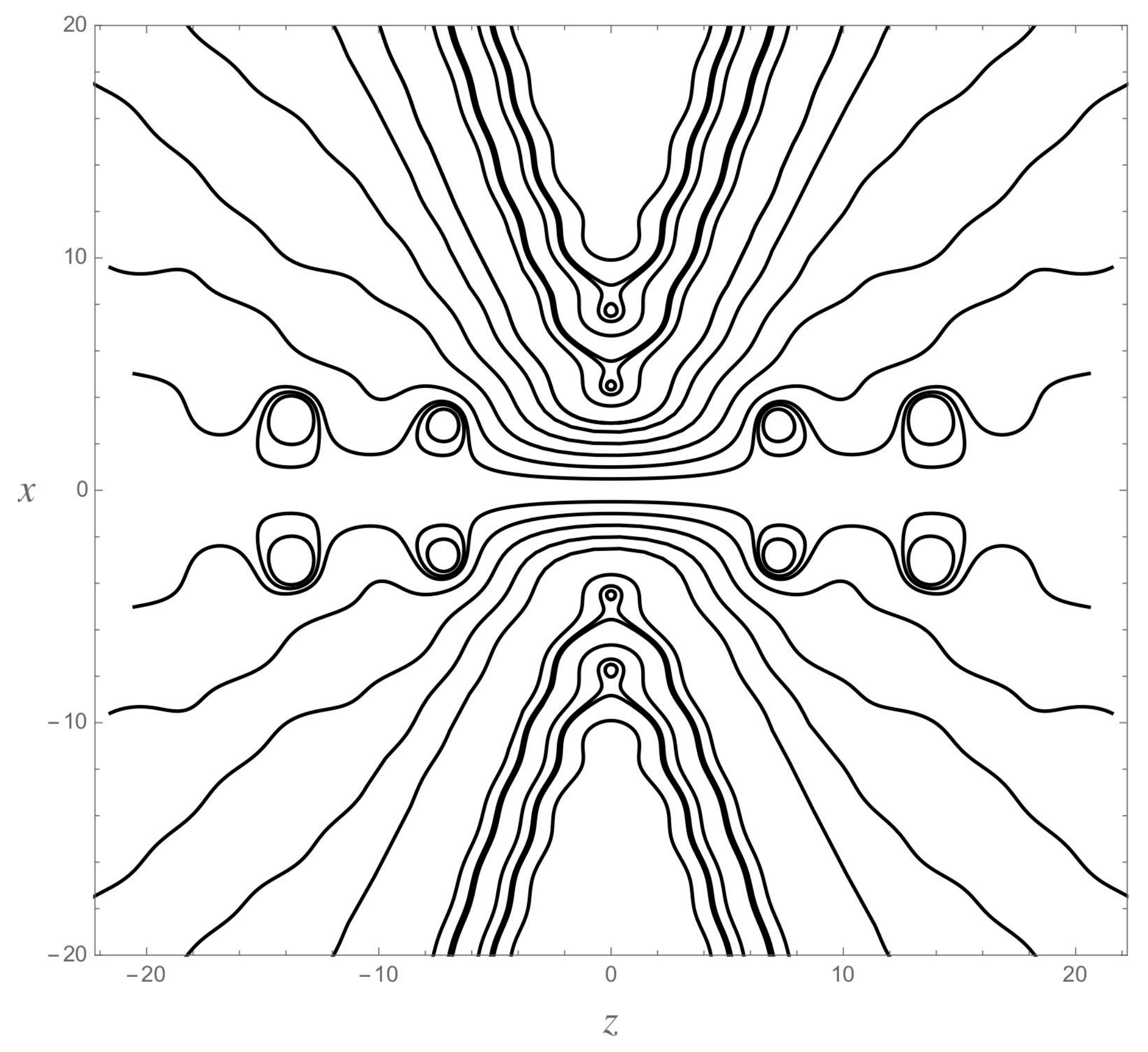}
\caption{ Integral curves of the time-averaged Poynting vector in the $xz$-plane for the radially polarized beam of Sec.~\ref{sec:rad}. Vortices are visible in some regions near the focal plane $z=0$ and also in some regions near the beam axis. The parameters are $\mu_0=c=k_0=\omega=p_0=1$. }
\label{fig:Savrad}
\end{center}
\end{figure}
\subsection{Radially/azimuthally polarized beam}   \label{sec:rad}
We first choose $\bi{A}_0(\bi{r})$ to lie in the $z$-direction with $z$-component proportional to the fundamental scalar beam (\ref{phibeam}). The real part of $\bi{A}_0(\bi{r})$ can then be written
\begin{equation}  \label{edstan}
\mathrm{Re}[\bi{A}_0(\bi{r})]= \frac{\mu_0\omega p_0 \sin(k_0 r)}{2\pi r} \boldsymbol{\hat{z}},
\end{equation}
which is simply the vector potential of an electric-dipole standing wave in Lorenz gauge, i.e.\  the retarded minus the advanced solution for electric-dipole radiation, where the constant $p_0$ is the amplitude of the oscillating dipole~\cite{jac}. Note that this standing wave is source free, like the scalar spherical standing wave (\ref{spherical}). From the familiar properties of electric-dipole radiation~\cite{jac}, one sees that the standing wave (\ref{edstan}) has $\bi{E}(\bi{r},t)$ lying in planes through the $z$-axis with $\bi{B}(\bi{r},t)$ in circles around the $z$-axis. The electric and magnetic fields of the beam have an additional contribution coming from the imaginary part of the scalar solution (\ref{phibeam}). But the resulting vector potential still has only a $z$-component and it depends spatially on the variables $r_\perp$ and $z$. As a result, one can see from (\ref{EBA}) that the fields in the beam have the properties stated above for the  standing wave (\ref{edstan}): the $\bi{B}$ field  lies in circles around the $z$-axis with the $\bi{E}$ field in planes through the $z$-axis. Since the beam propagates in the $z$-direction, we have a radially polarized beam, i.e.\ the $\bi{E}$ field has only radial and longitudinal ($z$-) components in cylindrical coordinates centred on the beam axis, while the $\bi{B}$ field only has an azimuthal ($\phi$-) component around the beam axis. By electromagnetic duality~\cite{jac}, we easily obtain an azimuthally polarized beam by swapping $\bi{E}$ and $\bi{B}$ (with the required constant factors). Figure~\ref{fig:fieldsrad} shows integral curves of the $\bi{E}$ and $\bi{B}$ fields in the radially polarized beam at four times over a half-period.  After half a period the directions of the field vectors are opposite to their original directions. It is notable how the structure of the $\bi{E}$ field changes significantly over a half-period. In particular, circular nodes of the $\bi{E}$ field appear and disappear in time. The $\bi{B}$ field also has circular nodes and some of them are fixed nodes that do not change in time (see below).

The Poynting vector $\bi{S}=\mu_0^{-1}\boldsymbol{E \! \times \! B}$ lies in planes through the beam axis and is rotationally symmetric around this axis. Figure~\ref{fig:Savrad} shows integral curves of the time-averaged Poynting vector $\langle \bi{S}\rangle$ in the $xz$-plane. Vortices are visible both in the focal-plane region and also in regions close to the beam axis. Because of the symmetry, all of these vortex regions are toroidally shaped and wrapped around the beam axis.

\begin{figure}[!htbp]
\begin{center} 
\includegraphics[width=14cm]{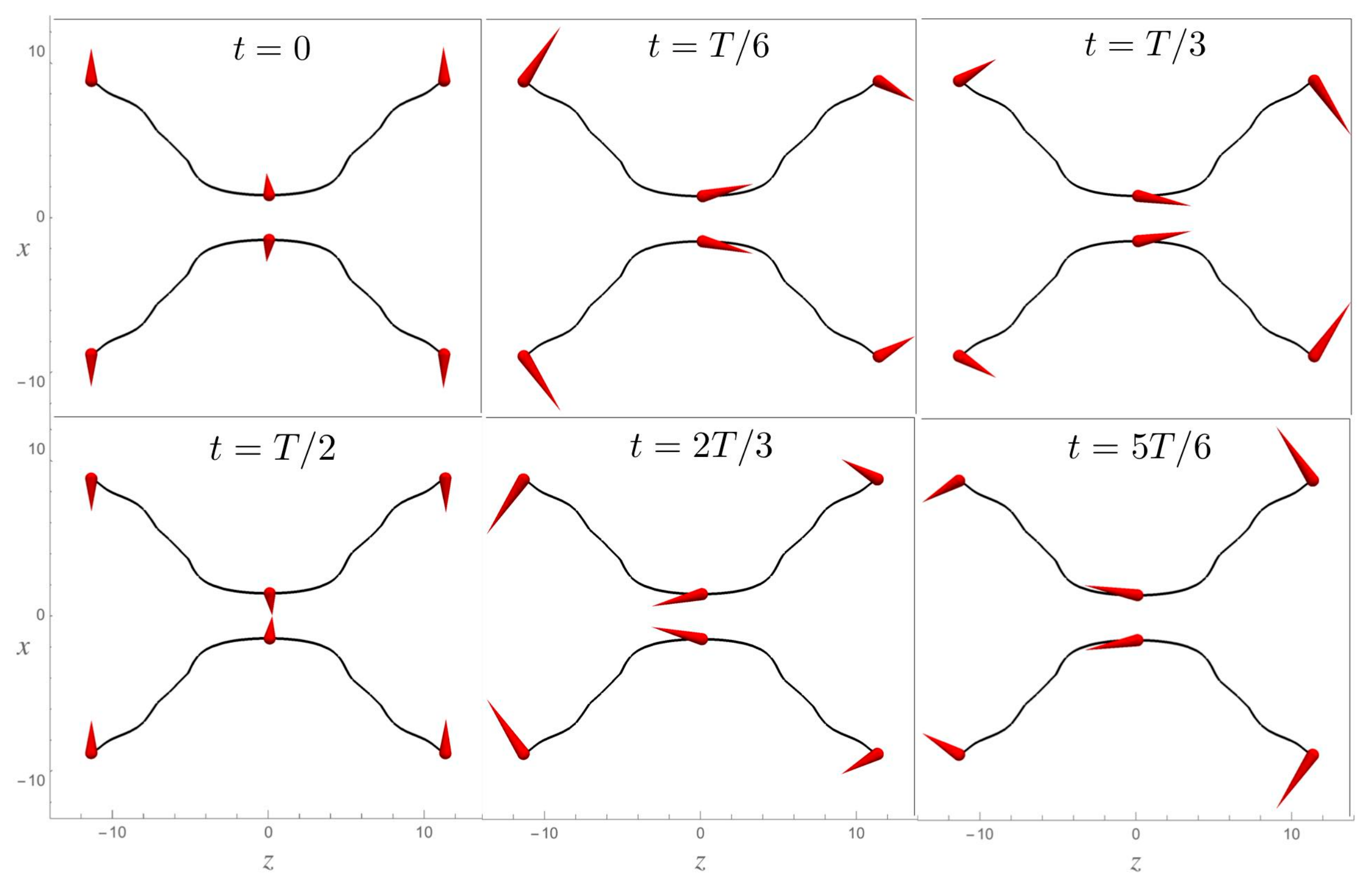}
\caption{Two integral curves of the time-averaged Poynting vector for the radially polarized beam of Sec.~\ref{sec:rad}, together with the electric field (red pointers) at six points over one period $T$. The field vector at the two points close to the focus $\bi{r}=0$ has been scaled by a factor of $6/70$ relative to the field at the other four points.  The magnetic field (not shown) at the six points oscillates perpendicular to the page. The parameters are $\mu_0=c=k_0=\omega=p_0=1$. }
\label{fig:SavErad}
\end{center}
\end{figure}

\begin{figure}[!htbp]
\begin{center} 
\includegraphics[width=13cm]{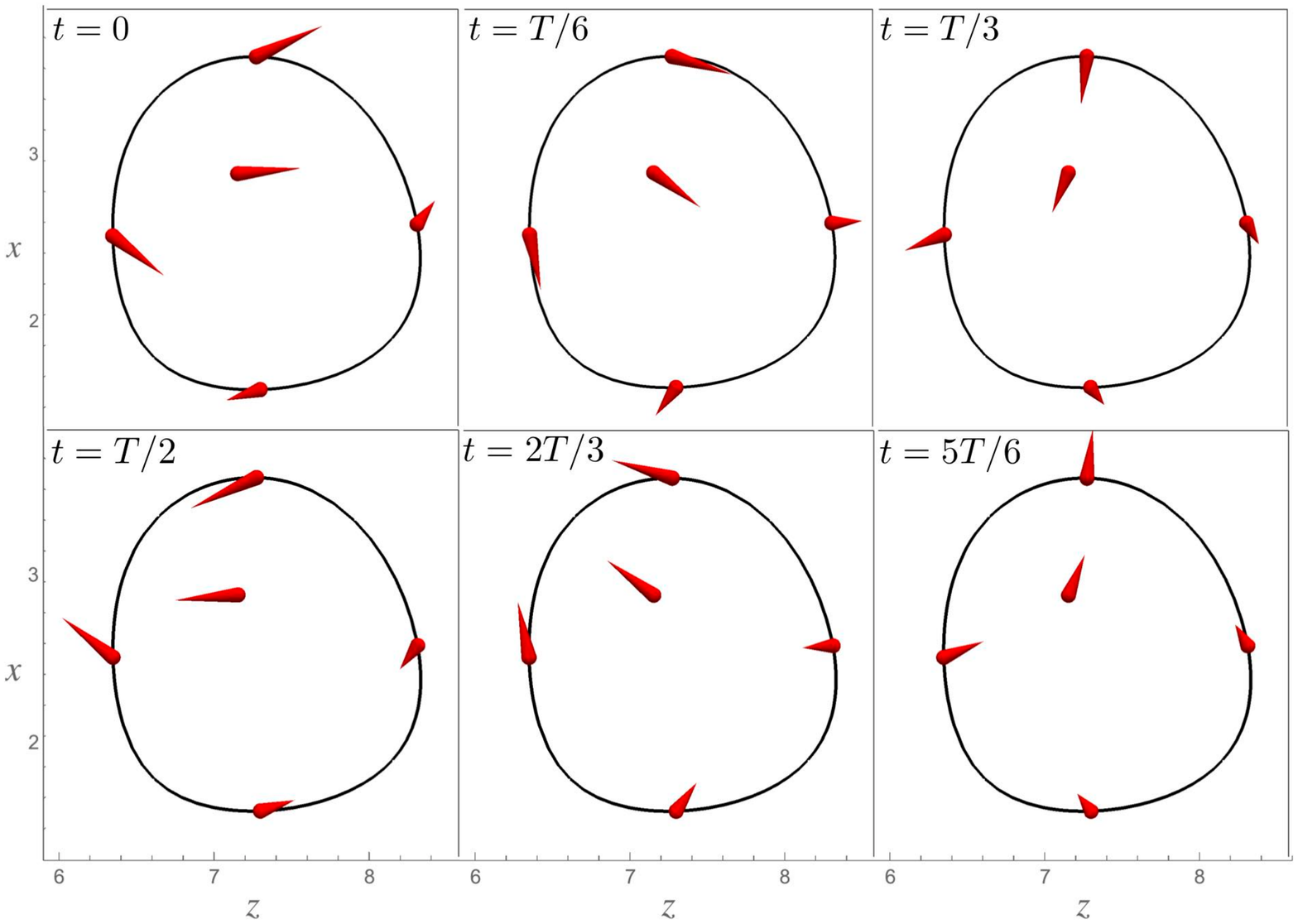}
\caption{A closed integral curve of the time-averaged Poynting vector in a vortex region of the radially polarized beam of Sec.~\ref{sec:rad}, together with the electric field (red pointers) at five points over one period $T$. The $\bi{E}$ vector at the point inside the loop lies at the centre of the vortex, at which the Poynting vector is zero. The vortex is a toroidal region around the beam axis and its centre is a circular node of the magnetic field. The parameters are $\mu_0=c=k_0=\omega=p_0=1$. }
\label{fig:vorErad}
\end{center}
\end{figure}

The Poynting vector itself (as opposed to its time average) only lies on the curves in  Figure~\ref{fig:Savrad} at one time during every half-period. This follows from the behaviour of the  $\bi{E}$ and $\bi{B}$ fields in time. Figure~\ref{fig:SavErad} shows the $\bi{E}$ field at six points in space during one period. The $\bi{B}$ field at these points (not shown) oscillates perpendicular to the page. Figure~\ref{fig:vorErad} is a similar plot showing the $\bi{E}$ field at five points in one of the vortex regions close to the beam axis. The $\bi{E}$ field at the point in Fig.~\ref{fig:vorErad} lying at the centre of the vortex is also shown. The vortex is a toroidal region around the beam axis and its centre is a circle on which the Poynting vector vanishes. As we see from the figure, the $\bi{E}$ field does not vanish at the vortex centre but the $\bi{B}$ field does vanish here (giving $\bi{S}=0$), and so the vortex centre is a fixed circular node of  $\bi{B}$. Elsewhere in Fig.~\ref{fig:vorErad} the $\bi{B}$ field (not shown) oscillates perpendicular to the page.

\begin{figure}[!htbp]
\begin{center} 
\includegraphics[width=8cm]{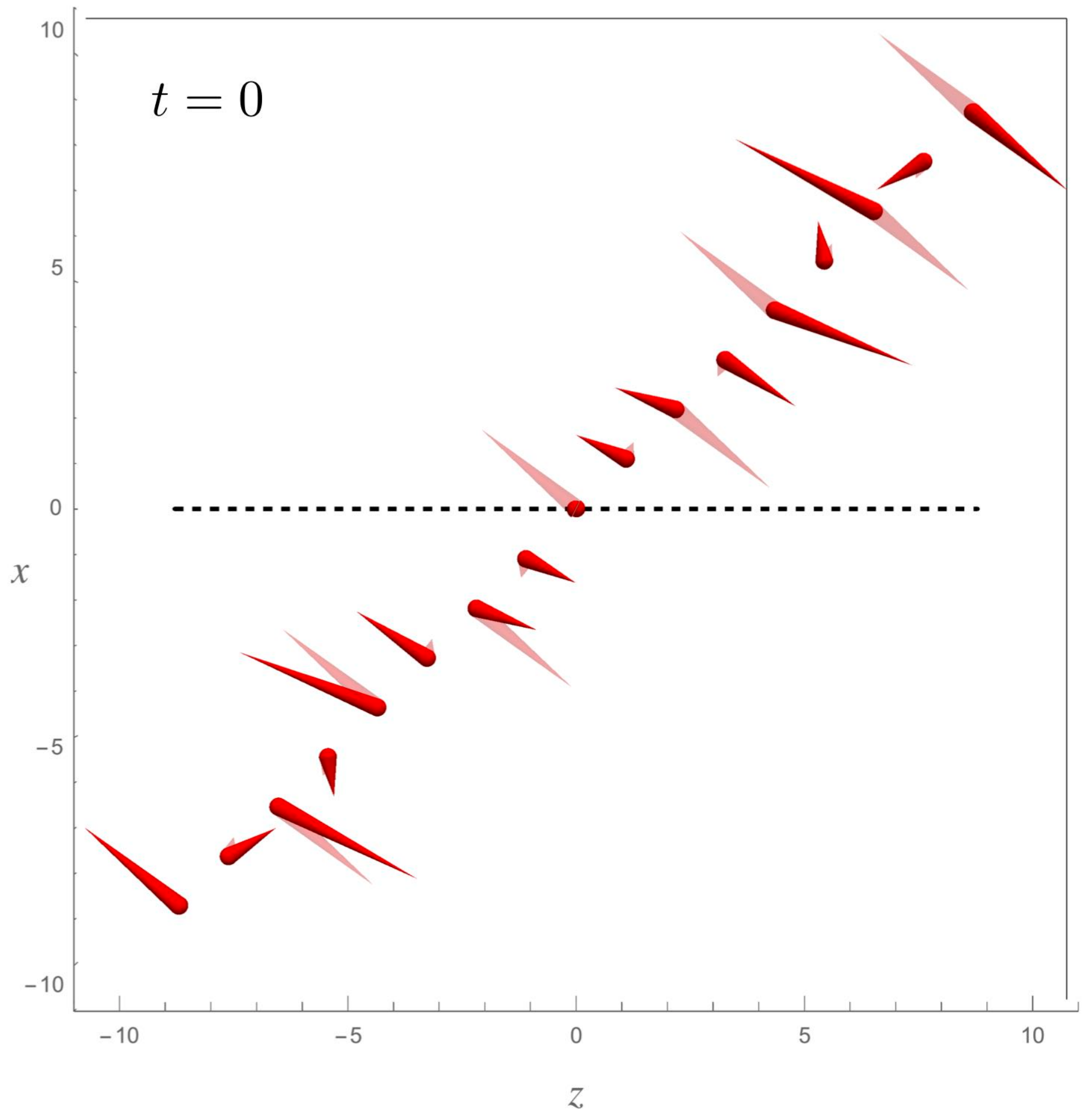}
\caption{Gouy phase shift in the radially polarized beam of Sec.~\ref{sec:rad}. The beam axis is the dotted line. The electric field (red pointers) is shown at points along a diagonal through the focus ($\bi{r}=0$) at time $t=0$, together with ``ghost" field vectors (pale red pointers) that show how the field would behave if the phase accumulation along the diagonal were that of a plane wave. At the end (top right) of the diagonal line the $\bi{E}$ field is opposite to the ghost field---this is the Gouy phase shift of $\pi$. The $\bi{E}$ field (red pointers) has been scaled by a factor proportional to the distance from the focus. The distance between the field vectors is $1/4$ of the free-space wavelength $2\pi/k_0$. The parameters are $\mu_0=c=k_0=\omega=p_0=1$. }
\label{fig:gouy}
\end{center}
\end{figure}

The beam exhibits the usual Gouy phase shift of $\pi$ through the focus~\cite{born}, as shown for the electric field in Fig.~\ref{fig:gouy} (the magnetic field of course also has a Gouy phase shift of $\pi$). We include this figure because of some recent discussion of the Gouy phase for radially polarized beams~\cite{pan13,kal16}. Once the electric field vectors are plotted as in Fig.~\ref{fig:gouy} it is clear that there is nothing unusual about the Gouy phase in this case. In~\cite{pan13,kal16} the component of the $\bi{E}$ field transverse to the beam direction is taken relative to a radial unit vector perpendicular to the beam axis. Note that this unit vector changes direction along a line through the beam axis. This change in direction of the radial unit vector is attributed to the $\bi{E}$ field in~\cite{pan13,kal16}, giving an extra sign change in the transverse component of $\bi{E}$ in addition to its Gouy phase shift. The same procedure applied to a linearly polarized plane wave propagating across  the $z$-axis would attribute a sign change (phase shift of $\pi$) to the transverse component ($\perp\boldsymbol{\hat{z}}$) of $\bi{E}$, although there is no focusing and no Gouy phase shift.

\subsection{Beam with angular momentum in the propagation direction}  \label{sec:radoam}
\begin{figure}[!htbp]
\begin{center} 
\includegraphics[width=15.5cm]{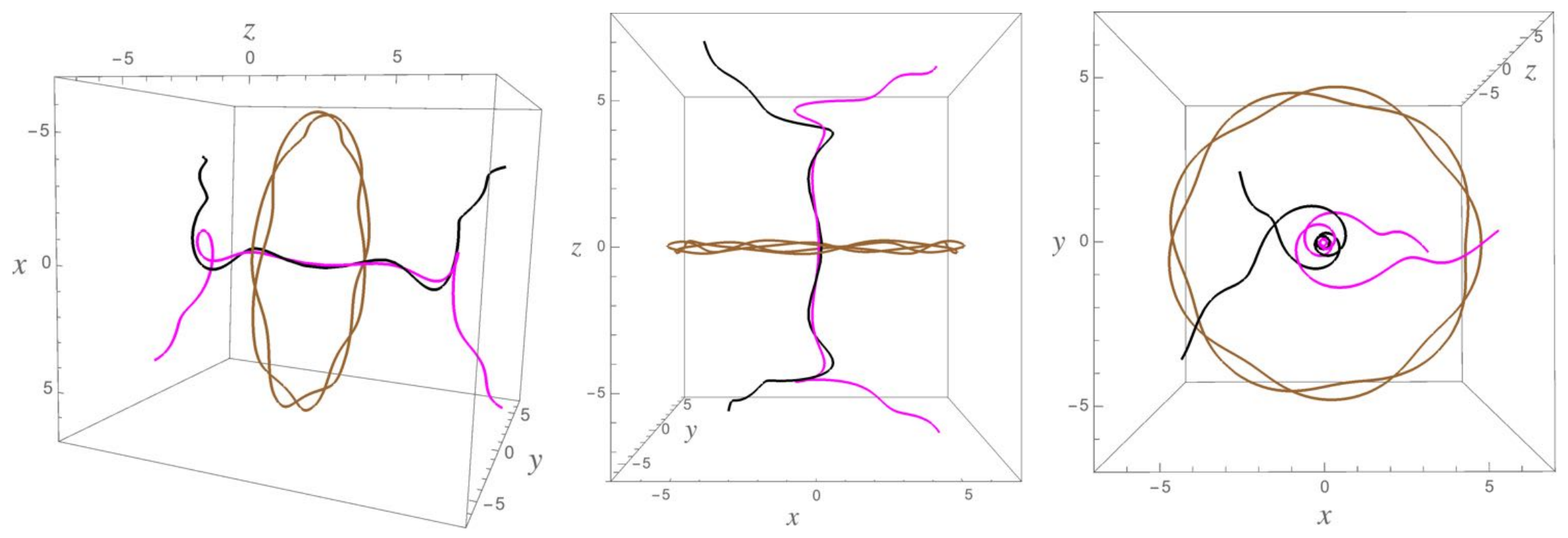}
\caption{Integral curves of the time-averaged Poynting vector for the beam obtained by acting with $\partial_x+\rmi \partial_y$ on the radially polarized beam of section~\ref{sec:rad}  Three views of the same integral curves are shown. The parameters are $\mu_0=c=k_0=\omega=p_0=1$. }
\label{fig:SavvecOAM}
\end{center}
\end{figure}

\begin{figure}[!htbp]
\begin{center} 
\includegraphics[width=15.5cm]{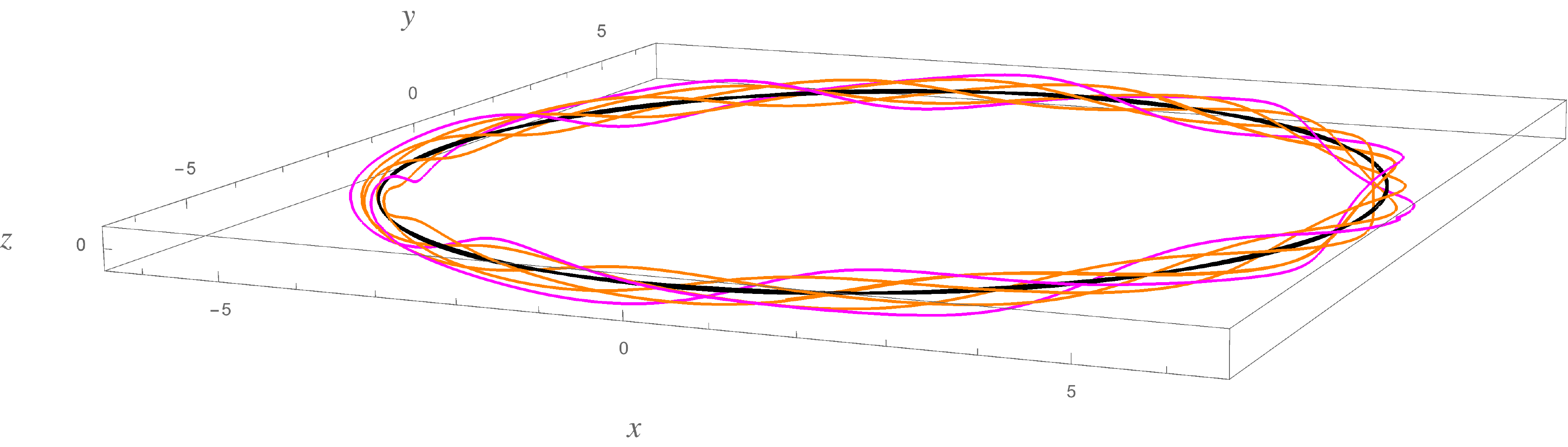}
\caption{Three closed loops of $\langle \bi{S}\rangle$ in the vortex region of Fig.~\ref{fig:SavvecOAM}. The circular loop of $\langle \bi{S}\rangle$ is the centre of the vortex. The other two loops wind around the circular centre. The number of windings increases for loops closer to the centre. The parameters are $\mu_0=c=k_0=\omega=p_0=1$. }
\label{fig:SavvecOAMvor}
\end{center}
\end{figure}

We obtain a light beam with  angular momentum in the propagation direction by acting on the radially polarized beam of the last subsection with $\partial_x+\rmi \partial_y$. Figure~\ref{fig:SavvecOAM} shows some integral curves of the time-averaged Poynting vector $\langle \bi{S}\rangle$ for the resulting beam. The angular momentum density is  $\varepsilon_0\boldsymbol{r\times}\!(\boldsymbol{E \! \times \! B})$, which is proportional to $\boldsymbol{r\times S}$, and the angular momentum density in the direction of propagation is clear from the fact that $\langle \bi{S}\rangle$ is not confined to planes through the beam axis. Also visible in Fig.~\ref{fig:SavvecOAM} is a vortex region encircling the beam axis. As in the scalar beam with OAM in Sec.~\ref{sec:scal}, the integral curves of $\langle \bi{S}\rangle$ in the vortex region form closed loops which wind around the central circle of the vortex. But there is a major difference from the scalar case. In the scalar OAM case the energy flow was zero on the central circle of the vortex, whereas in the electromagnetic example the central circle of the vortex   is in fact a loop of $\langle \bi{S}\rangle$. This can be seen in Fig.~\ref{fig:SavvecOAMvor}, which shows some closed loops of $\langle \bi{S}\rangle$ in the vortex region, including the circular loop of $\langle \bi{S}\rangle$ at the centre of the vortex. The winding of $\langle \bi{S}\rangle$ around the vortex centre shows that the beam has angular momentum about the centre of the vortex, i.e.\ in the vortex region there is angular momentum density transverse to the beam direction. Light beams that carry angular momentum transverse to the beam direction are discussed in~\cite{leu15}, but note that the same phenomenon occured for the scalar beam with OAM in Sec.~\ref{sec:scal} (see Fig.~\ref{fig:SavscalOAM}) so this effect does not require the field to have spin. 

\begin{figure}[!htbp]
\begin{center} 
\includegraphics[width=13cm]{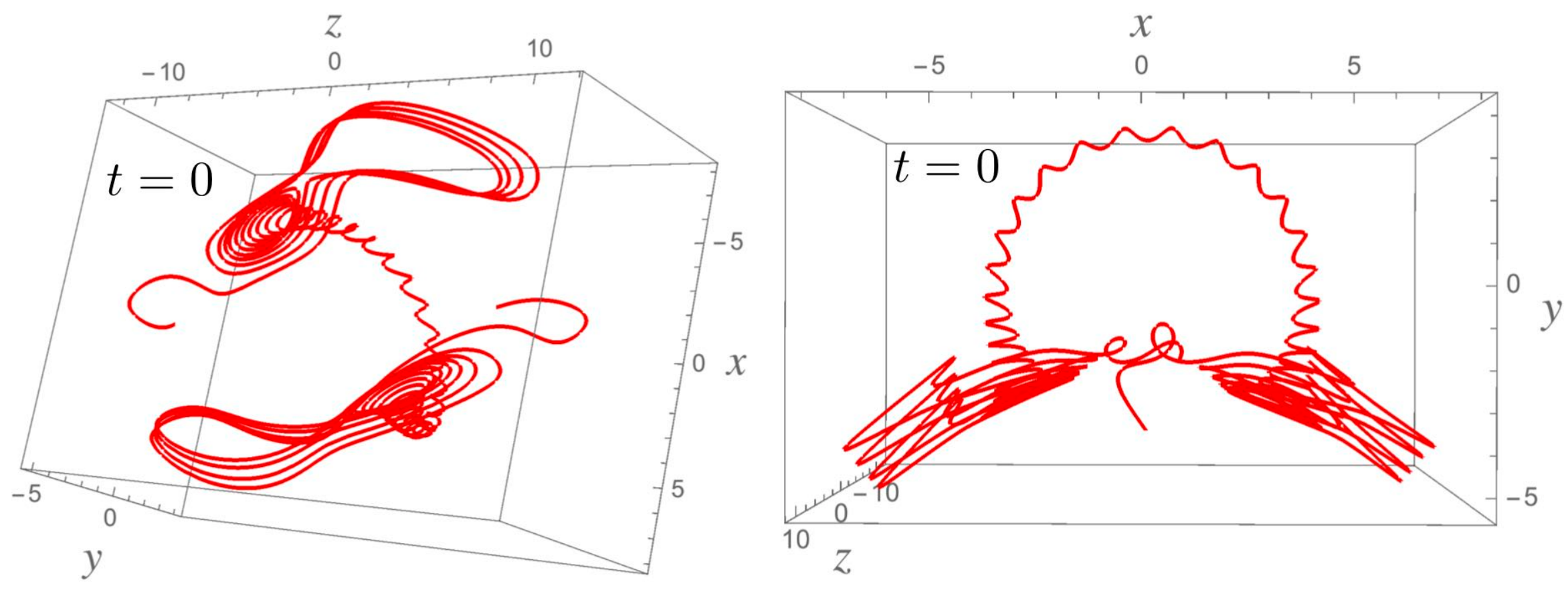}
\caption{Part of one integral curve of the electric field for the beam shown in Figs.~\ref{fig:SavvecOAM} and~\ref{fig:SavvecOAMvor}, at time $t=0$. Two views of the curve are shown.  The parameters are $\mu_0=c=k_0=\omega=p_0=1$. }
\label{fig:EvecOAM}
\end{center}
\end{figure}

The structure of the electric field in this beam is remarkably intricate, as is indeed the case for any beam in the class whose polarization cannot be simply described. Recall that in the radially polarized beam from which the beam in this subsection was generated, the  $\bi{E}$ field lines were in planes through the beam axis (see Fig.~\ref{fig:fieldsrad}).  Now the $\bi{E}$ field lines follow highly convoluted paths in space, which moreover change significantly over a half period. To give a small flavour of the behaviour, Fig.~\ref{fig:EvecOAM} shows part of one $\bi{E}$ field line at one time.

\subsection{Beam circularly polarized on its axis}   \label{sec:cir}

By choosing $\bi{A}_0(\bi{r})$ to have only an $x$-component that is proportional to the fundamental scalar beam (\ref{phibeam}), we obtain a beam that is linearly polarized in the $x$-direction on the beam axis. Specifically, if $\bi{A}_0(\bi{r})=\mu_0\omega p_0\phi(\bi{r}) \boldsymbol{\hat{x}}$, with $\phi(\bi{r})$ given by (\ref{phibeam}), the $\bi{E}$ field on the beam axis is in the $x$-direction. The  expressions for $\bi{E}$ and $\bi{B}$ on the beam axis ($z$-axis) are found as in the derivation of (\ref{scalzaxis}) above, and asymptotically we have
\begin{equation}  \label{fieldscirc}
\fl
\bi{E}_0(0,0,z) \sim \frac{\mu_0\omega^2 p_0}{4\pi z}\left( 2e^{\rmi k_0 z} -1 \right)  \boldsymbol{\hat{x}},
\qquad \bi{B}_0(0,0,z) \sim \frac{\mu_0\omega^2 p_0}{2\pi c z}e^{\rmi k_0 z} \boldsymbol{\hat{y}}, \qquad |z|\to\infty.
\end{equation}
As explained in the discussion of the radially polarized beam in Sec.~\ref{sec:rad}, the real parts of $\bi{E}_0(\bi{r})$ and $\bi{B}_0(\bi{r})$ correspond to an electric-dipole standing wave. We give no further details of this beam here, except to note that it does not carry angular momentum in the propagation direction (the Poynting vector lies in planes through the beam axis).

\begin{figure}[!htbp]
\begin{center} 
\includegraphics[width=15.5cm]{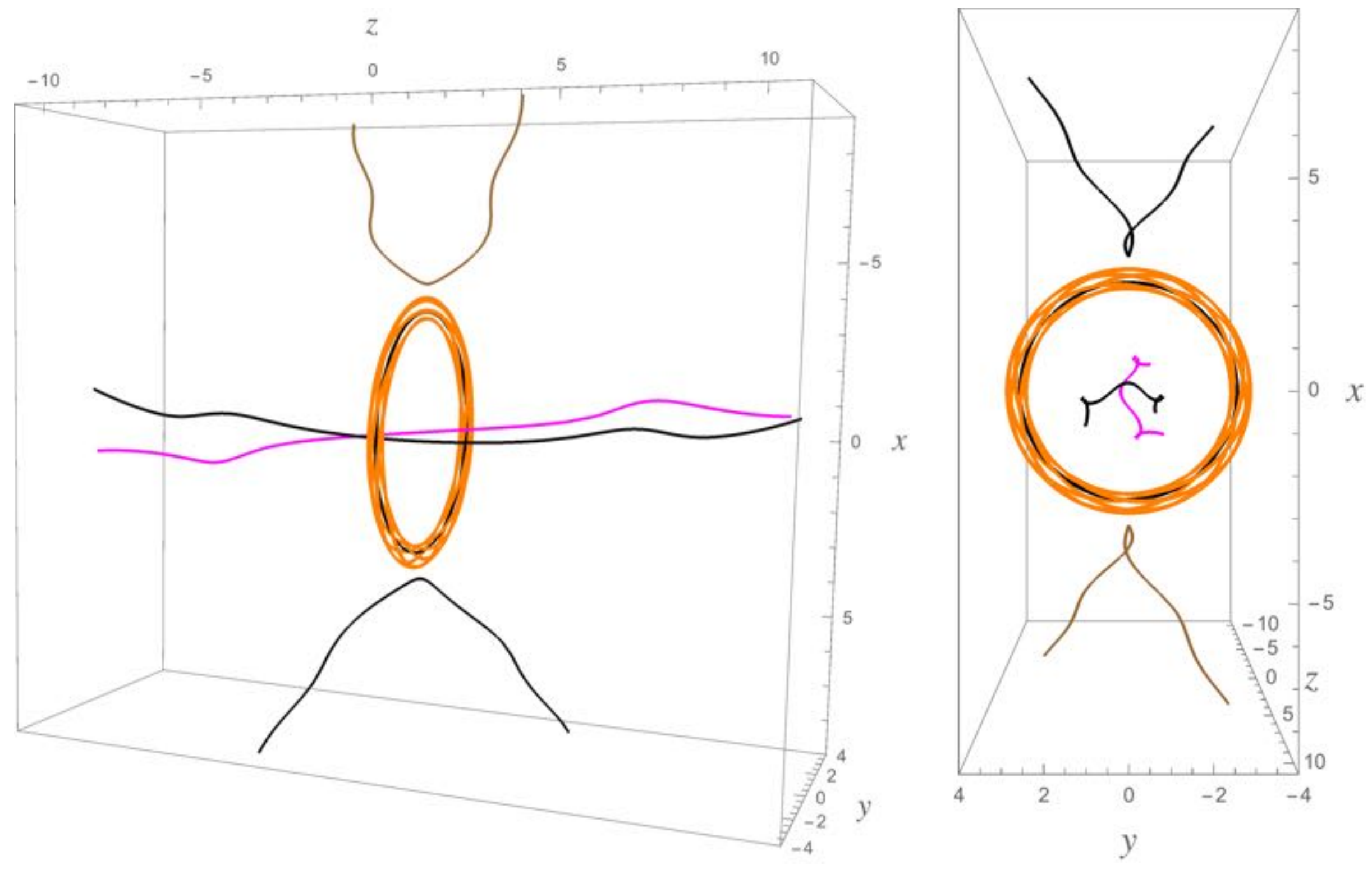}
\caption{Integral curves of $\langle \bi{S}\rangle$ in the beam described in Sec.~\ref{sec:cir}. Two views of the same integral curves are shown. Two of the curves are in a vortex region around the beam axis, one of them being the circular loop of $\langle \bi{S}\rangle$ at the centre of the vortex. The parameters are $\mu_0=c=k_0=\omega=p_0=1$. }
\label{fig:Savcir}
\end{center}
\end{figure}

Taking $\bi{A}_0(\bi{r})=\mu_0\omega p_0\phi(\bi{r}) \boldsymbol{\hat{y}}$, with $\phi(\bi{r})$ given by (\ref{phibeam}), we have a beam linearly polarized in the $y$-direction along the beam axis. Putting a factor of $\rmi$ in this $\bi{A}_0(\bi{r})$ and superposing it with the previous beam we generate a beam that is circularly polarized along the beam axis (left circularly polarized in the usual convention~\cite{jac}). The polarization away from the beam axis is not so simple to describe but because of the on-axis circular polarization we expect this beam to have angular momentum in the propagation direction (for right circular polarization we would expect the angular momentum to be opposite to the beam direction). The angular momentum density $\varepsilon_0\boldsymbol{r\times}\!(\boldsymbol{E \! \times \! B})$ is determined by the Poynting vector, and the angular momentum about the beam axis can be seen in the integral curves of the time-averaged Poynting vector in Fig.~\ref{fig:Savcir}. There are vortex regions in the beam, one of which is visible in the figure; the centre of the vortex is a circle of $\langle \bi{S}\rangle$ around the beam axis, as in  Fig.~\ref{fig:SavvecOAMvor}. 

It can be seen from Fig.~\ref{fig:Savcir} that the azimuthal component of $\langle \bi{S}\rangle$ transverse to the beam direction does not everywhere have the same sign. Although the time-averaged energy flux about the beam axis is predominantly anticlockwise for an observer facing the beam (as in the right-hand image in Fig.~\ref{fig:Savcir}), there are regions where it is clockwise. This should be clear from the right-hand plot in Fig.~\ref{fig:Savcir}, particularly in the two lines of $\langle \bi{S}\rangle$ furthest from the focus. In fact the same phenomenon occurs for the previous light beam with angular momentum in the propagation direction, described in Sec.~\ref{sec:radoam} (see the  right-hand plot in Fig.~\ref{fig:SavvecOAM}). This effect has also been seen in diffraction-theory results for the focusing of certain Laguerre-Gaussian beams with a lens~\cite{mon09}.

\subsection{Knotted field lines in beams}   \label{sec:knots}
In Sec.~\ref{sec:radoam} we briefly discussed the field lines in one of the beams. Unless the beam has a relatively simple polarization behaviour (e.g.\ the radial/azimuthal polarizations of Sec.~\ref{sec:rad}), it appears that the field lines always have a very intricate, indeed knotty, three-dimensional structure. The occurrence of knotted field lines in electromagnetic waves is studied in~\cite{irv08,ked13,cam18}, for example. Here we will plot some field lines for a beam in which there are some relatively simple and symmetrical field knots. Consider the  beam  $\bi{A}_0(\bi{r})=\mu_0\omega p_0\phi(\bi{r}) \boldsymbol{\hat{y}}$, with $\phi(\bi{r})$ given by (\ref{phibeam}). As noted in  Sec.~\ref{sec:rad}, the real part of the complex fields $\bi{E}_0(\bi{r})$ and $\bi{B}_0(\bi{r})$ for such a beam are those of an electric-dipole standing wave. Now perform a duality transformation $\bi{E}\to c \bi{B}$, $\bi{B}\to - \bi{E}/c$ on this beam; the result is a new beam for which the real parts of $\bi{E}_0(\bi{r})$ and $\bi{B}_0(\bi{r})$ are those of a magnetic-dipole standing wave. Form an equal superposition of these two beams but with a factor of $\rmi$ in the complex fields $\bi{E}_0(\bi{r})$ and $\bi{B}_0(\bi{r})$ of the first beam. The result is a beam for which the real parts of $\bi{E}_0(\bi{r})$ and $\bi{B}_0(\bi{r})$ are the standing wave (retarded minus advanced solutions) for an electric-dipole point source on top of a magnetic-dipole point source, both oriented in the  $\boldsymbol{\hat{y}}$ direction but out of phase with each other. The beam also contains the Hilbert transform in $z$ of this standing wave and propagates in the $z$ direction, as explained in Sec.~\ref{sec:standing}.

\begin{figure}[!htbp]
\begin{center} 
\includegraphics[width=11cm]{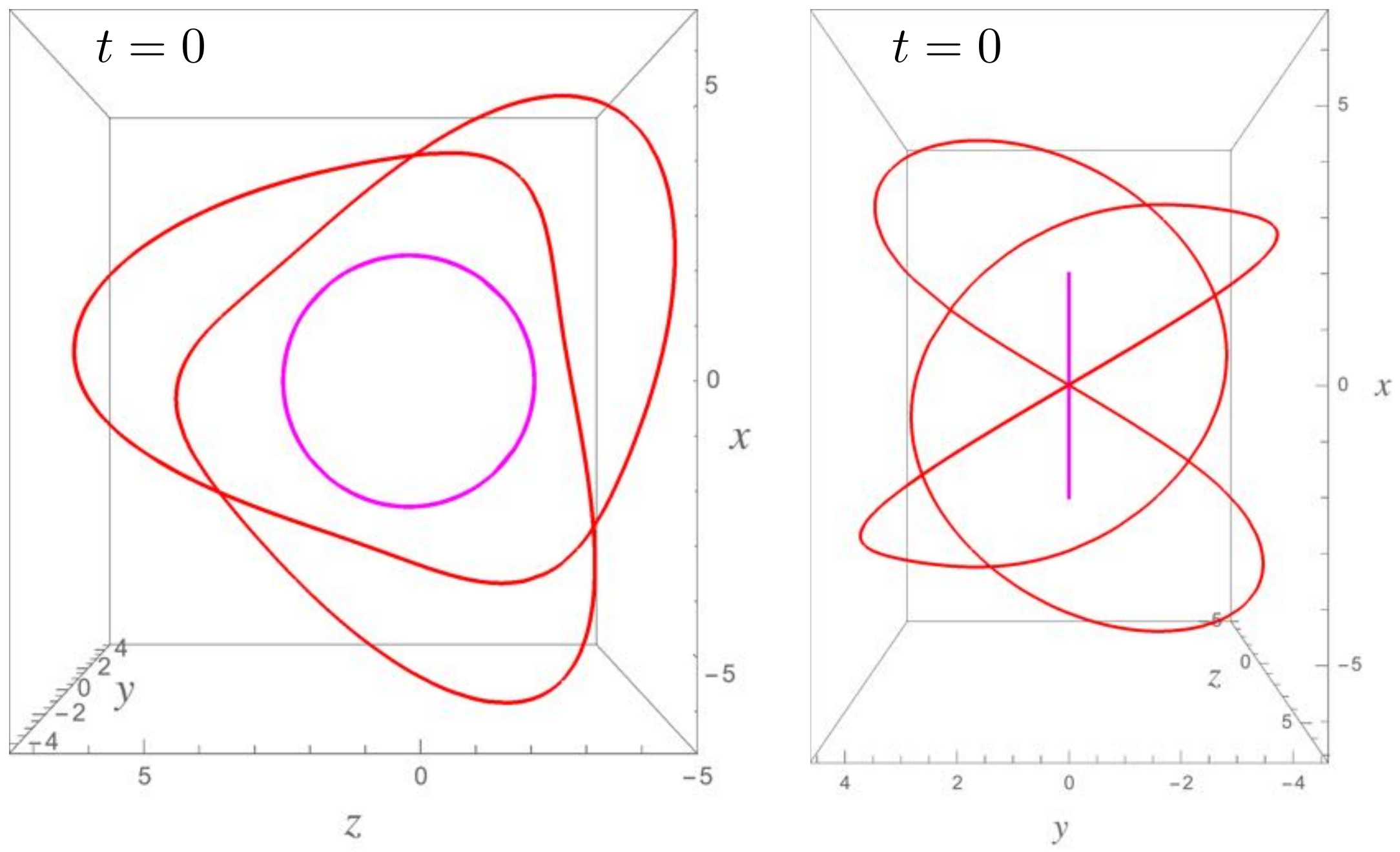}
\caption{ Two electric field lines in the light beam described in Sec.~\ref{sec:knots}, at time $t=0$. Two views of the same field lines are shown. One field line is a circle and the other is a trefoil knot. The parameters are $\mu_0=c=k_0=\omega=p_0=1$. }
\label{fig:knot1}
\end{center}
\end{figure}
\begin{figure}[!htbp]
\begin{center} 
\includegraphics[width=11cm]{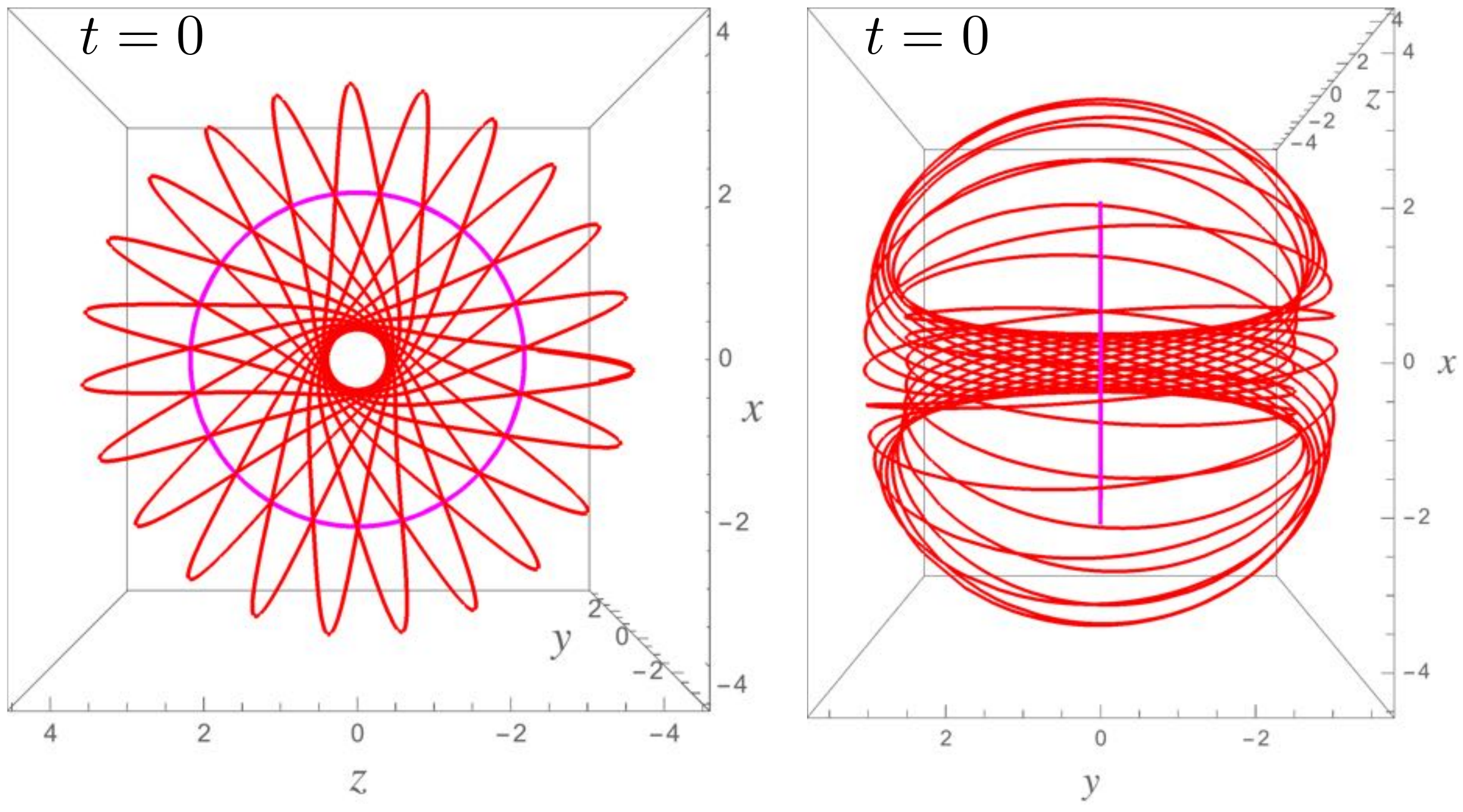}
\caption{Two electric field lines in the light beam described in Sec.~\ref{sec:knots}, at time $t=0$. Two views of the same field lines are shown. The circular field line is the same as in Fig.~\ref{fig:knot1}. The parameters are $\mu_0=c=k_0=\omega=p_0=1$. }
\label{fig:knot2}
\end{center}
\end{figure}
\begin{figure}[!htbp]
\begin{center} 
\includegraphics[width=11cm]{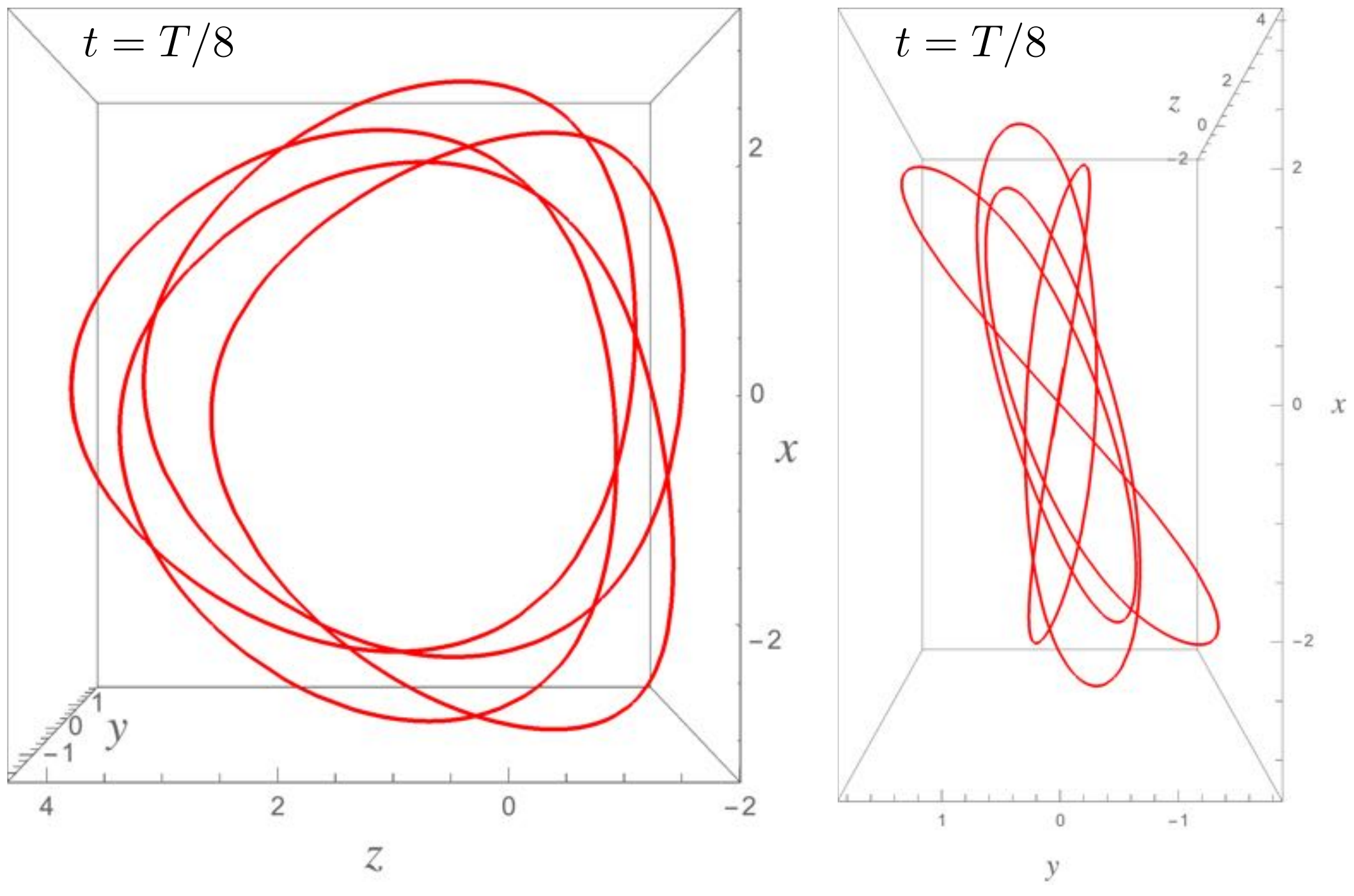}
\caption{ One electric field line in the light beam described in Sec.~\ref{sec:knots}, at time $t=T/8$, where $T$ is the period. The field line forms a knot with 15 crossings. The parameters are $\mu_0=c=k_0=\omega=p_0=1$. }
\label{fig:knot3}
\end{center}
\end{figure}

Figures~\ref{fig:knot1}--\ref{fig:knot3} show some electric field lines in the beam described in the last paragraph. At $t=0$ the beam coincides exactly with the standing wave based on out-of-phase electric and magnetic dipole sources, described above. This standing wave was recently considered in~\cite{cam18} and found to contain knotted electric field lines. We see this in the $t=0$  plots, Figs.~\ref{fig:knot1} and~\ref{fig:knot2}, for the beam. But the field lines in the beam change significantly over a half period, within which time the beam does not coincide with the simple standing wave. Nevertheless knotted field lines are still found in the beam at other times, as shown for $t=T/8$ in Fig.~\ref{fig:knot3} ($T$ is the wave period). We remark in passing that this beam carries angular momentum in the propagation direction.

\section{Further developments}
As shown in Sec~\ref{sec:standing},  every monochromatic beam can be decomposed into two standing waves, each proportional to a Hilbert transform of the other. Any standing wave can be used to generate a beam by finding its Hilbert-transform partner. The beams in this paper are of particular interest because we have exact expressions in $\bi{r}$-space for both standing wave components. This in turn was a result of choosing a particularly simple initial standing wave whose Hilbert transform could be evaluated exactly. But the procedure can be applied to any choice of standing wave. In almost all cases the Hilbert transform will have to be performed numerically, so exploring the resulting beams will require numerical evaluation of integrals at each point in space. Moreover an initial standing wave may be chosen that does not have an exact analytical expression in $\bi{r}$-space. There is one significant advantage to generating a beam from a standing wave, even when it does not give an analytical solution for the beam in $\bi{r}$-space. We saw in Sec.~\ref{sec:standing} that the field of the beam coincides with that of the original standing wave at two times during every period. We therefore obtain control over the shape of the beam if we can control the shape of the original standing wave. It is in fact easy to generate standing waves with any required shape, as we now describe.

\begin{figure}[!htbp]
\begin{center} 
\includegraphics[width=15.5cm]{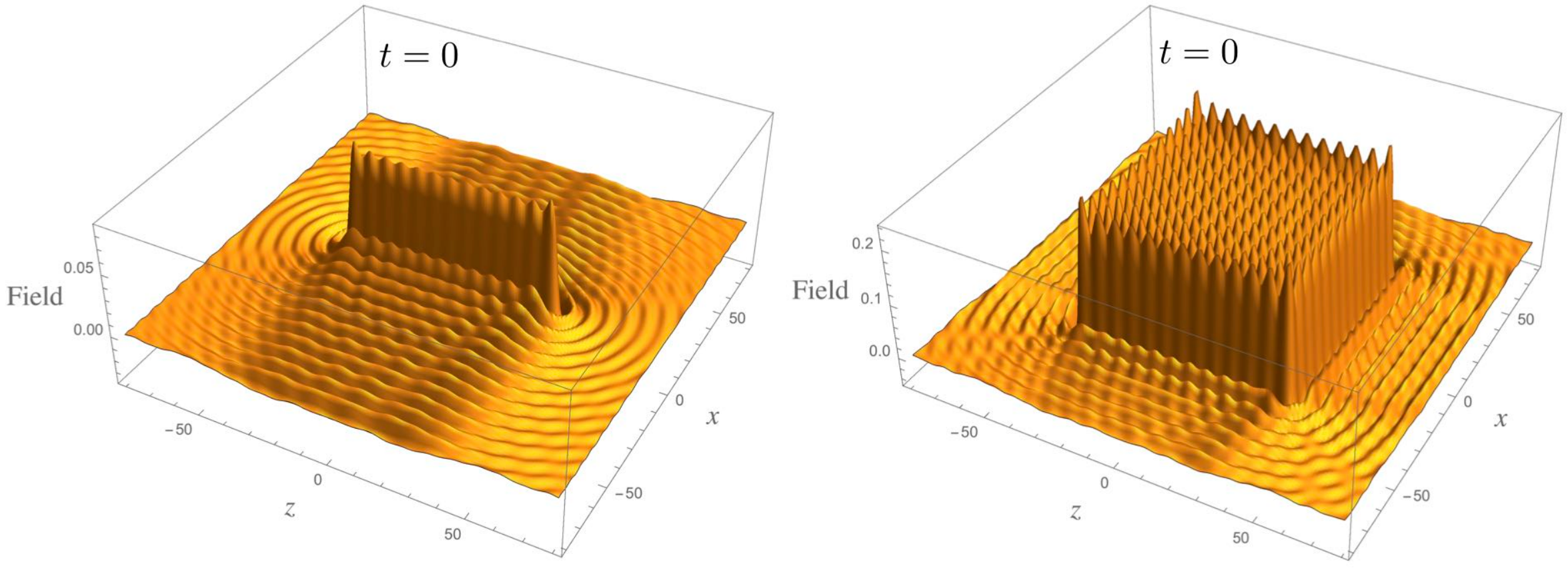}
\caption{Two (source-free) scalar standing waves. The waves are given by the retarded minus the advanced solutions for a uniform line-segment source on the $z$-axis  (left) and a uniform square source in the $xz$-plane (right). The line-segment source is of length $80/k_0$, which is also the length of the sides of the square source. The field has maximum positive value at the time shown ($t=0$). Beams are obtained from these standing waves $\phi_R(\bi{r})$ by adding $\rmi$ times the Hilbert transform with respect to any direction. The parameters are $c=k_0=\omega=1$. }
\label{fig:linesource}
\end{center}
\end{figure}

We noted in Secs.~\ref{sec:standing} and~\ref{sec:rad} that the original standing wave used for our beams is the retarded minus the advanced solution for a point source, essentially the retarded minus the advanced Green functions for the Helmholtz equation. Subtracting the advanced from the retarded solution for any monochromatic source gives a \emph{source-free} monochromatic wave, though not necessarily a standing wave. We can write the expression for this source-free wave by integrating up the source with the subtraction (retarded minus advanced) of the two Green functions. For the vector potential in the Lorenz gauge the expression is
\begin{equation}  \label{Ajgen}
 \bi{A}_0(\bi{r})=\frac{\rmi\mu_0}{2\pi}\int\rmd\bi{r'}\, \frac{\bi{j}_0(\bi{r'})\sin(k_0|\bi{r}-\bi{r'}|)}{|\bi{r}-\bi{r'}|},
\end{equation}
where $\mathrm{Re}[\bi{j}_0(\bi{r})e^{-\rmi\omega t}]$ is the arbitrary monochromatic source current density. If $\bi{j}_0(\bi{r})$ is real up to a complex constant factor, then (\ref{Ajgen}) is a (source-free) standing wave. For example, the standing wave (\ref{edstan}) is (\ref{Ajgen}) with the point electric-dipole source $\bi{j}_0(\bi{r})=-\rmi\omega p_0 \delta(\bi{r}) \boldsymbol{\hat{z}}$. By choosing $\bi{j}_0(\bi{r})$ to be real (up to a constant complex factor), and nonzero only inside some region $U\subset\mathds{R}^3$, we force the standing wave to have a central ``spot" of the same general shape as $U$. We illustrate this for a scalar wave in Fig.~\ref{fig:linesource}, in the cases when $U$ is a line segment or a square. The beam produced from the standing wave by the Hilbert-transform method will inherit the general shape of the standing wave, thus giving control over the shape of the focal region in the beam.

Note that the size of $U$ can be chosen independently of the ``free-space" wavelength $2\pi/k_0$. There is a diffraction limit on the smallness of the resulting spot size, regardless of how small $U$ is taken, and this limit is encountered in our class of beams, for which $U$ is a point. But the size of $U$ may be chosen arbitrarily large, and this will give the resulting beam a structure on a length scale arbitrarily larger than the free-space wavelength (see Fig.~\ref{fig:linesource}, for example). It has been pointed out that monochromatic beams can have structures on length scales much smaller than the free-space wavelength (and structures on time scales much smaller than the wave period)~\cite{kar98,ber98}. These small-scale structures of beams are visible in vortex regions~\cite{kar98,ber98} and they also occur in the field lines of the light beams above, although we have not pursued this here. Perhaps less often discussed is the construction of beams that also have very large length-scale structures, a phenomenon that may prove useful for optical trapping purposes.

Finally we note that one can use (\ref{Ajgen}) to write down a beam directly, by choosing $\bi{j}_0(\bi{r})$ to have only positive $k_z$ Fourier components. In  $\bi{k}$-space this is no different from constructing a beam by populating the $k_z>0$ hemisphere of the $k_0$-sphere, which was discussed in the Introduction. We see this by writing the right-hand side of (\ref{Ajgen}) in terms of Fourier components:
\begin{equation}   \label{AjgenFouier}
 \bi{A}_0(\bi{r})= \frac{\rmi\mu_0}{(2\pi)^3} \int \rmd\bi{k} \, \frac{\pi}{k_0} \delta(k-k_0)  \bi{j}_0(\bi{k}) e^{\rmi\bi{k\cdot r}}.
\end{equation}
Starting directly in $\bi{k}$-space however makes it difficult to control the shape of the beam. 
The control over the beam shape we have discussed above requires us to choose $\bi{j}_0(\bi{r})$ in $\bi{r}$-space, which will generally give $\bi{j}_0(\bi{k})\neq 0$, for $k_z<0$. We must then remove negative $k_z$ components from the resulting wave to obtain a beam, and this is what the Hilbert-transform prescription achieves. In terms of $\bi{k}$-space integrals the procedure is to choose $\bi{j}_0(\bi{r})$ in $\bi{r}$-space with a desired shape, then insert its Fourier transform into (\ref{AjgenFouier}) and evaluate the $k_z$ integral with lower limit 0. 

\section{Conclusions}
The qualitative features of the class of beams presented here will be found in other beams with strong focusing.  For example, the occurrence of vortices in focused waves has been found to be quite generic~\cite{kar98,ber98,nye98,rub17,and17}. This is one reason why the class of beams discussed here is worth studying, even though it corresponds to an experimentally unrealistic case of maximal focusing. For theoretical studies, the fact that we have exact  solutions in $\bi{r}$-space confers an advantage, as these are easy to plot and explore.  By means of the exact solutions, the highly intricate energy and field structure of strongly focusing light beams, with various polarization properties, can be examined with comparative ease. Once interesting features are identified, their behaviour for less extreme focusing can be investigated by removing plane-wave components and evaluating the beam by Fourier integrals. It is also possible to study the exact solutions analytically, for example by expanding the solutions around interesting features, but we have not pursued this here. The results given above are simply a brief visual overview of some of the beams.

Further study of exact light-beam solutions may offer insights into how the structure of the beam is related to that of its plane-wave components. For example, although all the plane-wave components in a beam propagate in the beam direction, the energy in some regions of the beam can flow in the opposite direction, and not just in vortex regions as shown in Sec.~\ref{sec:scal}. Better understanding of this phenomenon will help in the design of interesting new beams, with possible application to tractor beams~\cite{lee10,brz13,shv14}. The structure of the field lines in electromagnetic waves  is also highly nontrivial and of increasing interest~\cite{irv08,ked13,cam18}. Knotted field lines occur in the class of electromagnetic beams described here, and the field lines are in general highly intricate unless the beam has a particularly simple polarization structure (such as the radial/azimuthal polarization of Sec.~\ref{sec:rad}). There is scope for general methods that can characterise the field structure and elucidate the factors controlling it.

\section*{Acknowledgements}
I thank S.A.R.\ Horsley and D.B.\ Phillips for helpful discussions. I also thank a Referee for bringing Ref.~\cite{lek16} to my attention.


\end{document}